\documentclass[twocolumn,tighten,times,useAMS,usenatbib]{aastex631}

\newcommand{\jwst}{{\it JWST}}

\usepackage{epstopdf,soul}
\usepackage{color}
\usepackage{multirow}
\newcommand{\jl}{{SN~2010jl}}

\graphicspath{{./}{Figures/}}
\shorttitle{JWST Observations of Dust in SN~2010jl}
\shortauthors{Smith et al.}

\begin{document}
\title{JWST Spectra Indicate a Large Mass of Post-Shock Dust Formed by SN~2010jl}

\author[0000-0001-5510-2424]{Nathan Smith}
\affil{Steward Observatory, University of Arizona, 933 N. Cherry Avenue, Tucson, AZ 85721, USA}

\author[0000-0002-9301-5302]{Melissa Shahbandeh}
\affil{Department of Physics and Astronomy, Johns Hopkins University, Baltimore, MD 21218, USA}
\affil{Space Telescope Science Institute, 3700 San Martin Drive, Baltimore, MD 21218, USA}

\author[0000-0003-2238-1572]{Ori D.\ Fox}
\affil{Space Telescope Science Institute, 3700 San Martin Drive, Baltimore, MD 21218, USA}


\author[0000-0001-8385-3727]{Thomas Moore}
\affil{Space Telescope Science Institute, 3700 San Martin Drive, Baltimore, MD 21218, USA}

\author[0000-0003-0123-0062]{Jennifer E.\ Andrews}
\affil{Gemini Observatory, 670 N. Aohoku Place, Hilo, HI 96720, USA}

\author[0000-0001-5955-2502]{Thomas G.\ Brink}
\affil{Department of Astronomy, University of California, Berkeley, CA 94720-3411, USA}

\author[0000-0001-8033-1181]{Eli Dwek}
\affil{Observational Cosmology Lab, NASA Goddard Space Flight Center, Mail Code 665, Greenbelt, MD 20771, USA}

\author[0000-0003-0209-674X]{Michael Engesser}
\affil{Space Telescope Science Institute, 3700 San Martin Drive, Baltimore, MD 21218, USA}

\author[0000-0003-3460-0103]{Alexei V.\ Filippenko}
\affil{Department of Astronomy, University of California, Berkeley, CA 94720-3411, USA}

\author[0000-0002-9915-1372]{Bryony Nickson}
\affil{Space Telescope Science Institute, 3700 San Martin Drive, Baltimore, MD 21218, USA}

\author[0000-0001-7380-3144]{Tea Temim}
\affil{Department of Astrophysical Sciences, Princeton University, 4 Ivy Lane, Princeton, NJ 08544, USA}

\author[0000-0002-6535-8500]{Yi Yang}
\affil{Department of Astronomy, University of California, Berkeley, CA 94720-3411, USA}
\affil{Department of Physics, Tsinghua University, Beijing 100084, China}

\author[0000-0002-2636-6508]{WeiKang Zheng}
\affil{Department of Astronomy, University of California, Berkeley, CA 94720-3411, USA}


\author[0000-0002-5221-7557]{Chris Ashall}
\affil{Institute for Astronomy, University of Hawai’i at Manoa, 2680 Woodlawn Dr., Hawai'i, HI 96822, USA}

\author[0009-0004-7268-7283]{Raphael Baer-Way}
\affil{Department of Astronomy, University of Virginia,  Charlottesville VA 22904-4325, USA}
\affil{National Radio Astronomy Observatory, 520.0 Edgemont Rd, Charlottesville VA 22903, USA}

\author[0000-0002-0141-7436]{Geoffrey C.\ Clayton}
\affil{Space Science Institute, 4765 Walnut St., Suite B Boulder, CO 80301, USA}

\author[0000-0001-5754-4007]{Jacob E. Jencson}
\affil{IPAC, MC 100-22, Caltech, 1200 E. California Blvd., Pasadena, CA 91125}

\author[0000-0001-5975-290X]{Joel Johansson}
\affil{Department of Physics, Oskar Klein Centre, Stockholm University, SE-106 91, Stockholm, Sweden}

\author[0009-0003-8380-4003]{Zachary G.\ Lane}
\affil{School of Physical and Chemical Sciences — Te Kura Matū, University of Canterbury, Private Bag 4800, Christchurch 8140, Aotearoa, New Zealand}

\author[0000-0002-0763-3885]{Dan Milisavljevic}
\affil{Purdue University, Department of Physics and Astronomy, 525 Northwestern Ave, West Lafayette, IN 47907, USA}

\author[0000-0002-4410-5387]{Armin Rest}
\affil{Space Telescope Science Institute, 3700 San Martin Drive, Baltimore, MD 21218, USA}
\affil{Department of Physics and Astronomy, The Johns Hopkins University, Baltimore, MD 21218, USA}

\author[0000-0002-9820-679X]{Arkaprabha Sarangi}
\affil{Indian Institute of Astrophysics, 100 Feet Rd, Koramangala, Bengaluru, Karnataka 560034, India}

\author[0000-0003-4610-1117]{Tam\'as Szalai}
\affil{Department of Experimental Physics, Institute of Physics, University of Szeged, D{\'o}m t{\'e}r 9, 6720 Szeged, Hungary}
\affil{MTA-ELTE Lend\"ulet "Momentum" Milky Way Research Group, Szent Imre H. st. 112, 9700 Szombathely, Hungary}

\author[0000-0001-9038-9950]{Schuyler D.\ Van Dyk}
\affil{IPAC, MC 100-22, Caltech, 1200 E. California Blvd., Pasadena, CA 91125}

\author[0000-0003-2063-381X]{Brian J. Williams}
\affil{NASA Goddard Space Flight Center, Code 662, Greenbelt, MD 20771, USA}

\begin{abstract}
We present new {\it James Webb Space Telescope} ({\it JWST}) mid-infrared (MIR) spectra and ground-based optical spectra of the lingering source at the position of supernova (SN)~2010jl, which was a relatively nearby superluminous Type~IIn supernova (SLSN~IIn) having strong interaction with circumstellar material (CSM). Early-time data showed evidence of dust, interpreted as either pre-existing CSM dust, or as newly formed dust in the SN ejecta and post-shock region.  At 13~yr post explosion, {\it JWST} reveals a strong MIR excess from warm dust, with broad features at 10--15 $\mu$m.  Our analysis reveals a minimum dust mass of $>$0.11 $M_{\odot}$, and a more likely value of 0.2 $M_{\odot}$ or more because the dust is optically thick.  This is among the largest masses of SN-produced dust yet measured without far-IR/submm data, and greatly exceeds SN~2010jl's dust mass inferred around 2--3~yr post-explosion.  Ground-based optical spectra confirm the presence of a young massive cluster at the SN position, and confirm that blueshifted line profiles persisting until the latest epochs arise from dust formed in the post-shock region.  The warmest dust emitting in the MIR is likely to be the same post-shock dust causing the blueshift.   The {\it JWST} spectrum also reveals silicate absorption, which may arise from cool SN ejecta dust along the line of sight to the receding shock.   The large mass of post-shock dust in SN~2010jl suggests that strong CSM interaction promotes efficient dust production, where the new post-shock dust will survive. If strongly interacting SNe are common in the early Universe, this may contribute significantly to dust seen in infant  galaxies.
\end{abstract}


\section{INTRODUCTION}\label{sec:intro}

As a result of both stellar and explosive nucleosynthesis \citep{arnettbook}, core-collapse supernovae (CCSNe) enrich the  interstellar medium (ISM) of galaxies with metals, and the refractory elements that they produce may also make them important sources of dust grains.  Significant amounts of dust observed in high-redshift galaxies \citep{bertoldi03,dunne03b,laporte17,sm23,langeroodi24,carniani24} may indicate that CCSNe are especially important dust producers in the early Universe, when sources of dust from evolved lower-mass stars (asymptotic giant branch (AGB) stars, novae, etc.) did not have enough time to contribute.

Far-infrared (IR) observations of dust in nearby SN remnants such as SN~1987A and Cas~A have revealed large masses of dust, of the order of 0.1--1 $M_{\odot}$ \citep{dunne03a,matsuura11,boccio16}, that are much larger than dust masses usually estimated from IR observations obtained shortly after an SN explosion. For example, in the case of SN~1987A, the huge mass of dust (0.4--0.8 $M_{\odot}$) inferred from far-IR observations decades after explosion \citep{matsuura11,matsuura15,indebetouw14} was much larger than the dust mass of 10$^{-4}$ to 10$^{-3}$ $M_{\odot}$ inferred from mid-IR (MIR) emission observed a few hundred days after explosion \citep{gn87,gn89,wooden93}.  Large dust masses exceeding 0.1 $M_{\odot}$ have also been inferred via line profile asymmetries in decades-old SNe such as SNe~1970G, 1979C, 1980K
and 1996cr \citep{nicu22}. This difference between the dust directly observed in fading SN events and the larger mass in older SNe or remnants may arise because the {\it inferred} dust mass continues to grow over time, 
dust clumps may be optically thick at early times \citep{dwek19}, or  shorter wavelength observations are only sensitive to the hottest dust which usually represents a small fraction of the total dust mass. Regardless, mid-IR to far-IR wavelengths are needed to see the cooler dust accumulating over time in cooling SN ejecta. Indeed, recent {\it James Webb Space Telescope} (\jwst) Mid-Infrared Instrument (MIRI) observations of the Type~IIP SN~2004et and Type~IIn SN~1995N uncovered some of the largest newly formed ejecta dust masses in extragalactic SNe besides SN~1987A \citep{shahbandeh23,clayton25}. 

SNe powered by interaction with circumstellar material (CSM), usually observed as Types IIn or Ibn (see \citealt{smith17} for a review of interacting SNe), represent an interesting case because they may form new dust in the post-shock zone \citep{smith08jc,smith12,ss22};  moreover, this dust may be kept warm by the ongoing CSM interaction shock \citep{smith08jc,fox10,fox11, fox13}, making this dust detectable in the MIR for many years longer than the MIR emission of non-interacting CCSNe.  Dust forming in these interacting SNe is of particular interest to the ISM dust budget, because dust that forms in the cooling post-shock layers is already behind the shock.  It will therefore not get destroyed by the forward or reverse shocks, and is more likely to survive to mix with the ISM.  This potential destruction is a significant uncertainty for dust formed in SN ejecta or the CSM dust formed during pre-SN mass-loss phases \citep{bs07,micelotta16,bc16}.   Thus, even though interacting SNe are rarer than non-interacting SNe ($< 10$\% of all CCSNe; see \citealt{smith11frac,pessi25}), their dust production makes them unique and potentially important in this regard. Moreover, owing to high densities and rapid cooling in the CSM interaction layers, this class of SN may form dust sooner after explosion than normal SNe \citep{smith08jc}.

While dust formation in SNe is expected to produce MIR excess emission, the interpretation of this IR emission alone can be ambiguous, since pre-existing dust in the CSM that becomes heated by the SN radiation (i.e., an ``IR echo") can also produce MIR emission. Generally, corroborating evidence of a faster decline in the light curve or a blueshift of emission lines --- both caused by increased extinction from new dust within the SN --- is needed to be confident that an event is actually producing new dust, as opposed to heating pre-existing dust. In the particular case of interacting SNe, one can diagnose if the newly formed dust is in the SN ejecta or in the post-shock zones by examining the blueshift in either the broad ($\sim 10^4$ km s$^{-1}$) emission-line components that trace the freely expanding SN ejecta, or the intermediate-width (IW; $\sim 1000$ km s$^{-1}$) components emitted by post-shock gas in the cold dense shell (CDS).  

The first clear case of an interacting SN that showed all three
signatures  of dust formation (MIR emission, increased fading rate, blueshifted line profiles) was the Type~Ibn SN~2006jc, where the line-profile
evolution of IW He~{\sc i} lines indicated that new
dust formed rapidly (by day 50) in the post-shock CDS \citep{smith08jc}.  This blueshift
coincided in time with an IR excess and fading in the continuum, but
also with a burst of X-ray emission and He~{\sc ii} $\lambda$4686 emission
\citep{immler08,smith08jc,dicarlo08}, indicating that the forward shock
encountering a dense shell is what triggered the rapid onset of dust
formation only 50--100 days after discovery.  The rapid post-shock dust
formation may be analogous to what happens in dust-forming
colliding-wind binaries like the carbon-rich WC+OB system WR~140
\citep{hgg79,williams90,monnier02} and $\eta$~Car
\citep{smith10eta}.  Similar blueshifted profiles have been seen in a
number of H-rich interacting SNe, including SN~2005ip \citep{smith09,fox10,smith17ip,fox20,bevan19}, SN~2006tf \citep{smith08tf}, SN~2007rt \citep{trundle09}, SN~2007od \citep{andrews10}, SN~2010bt \citep{eliasrosa18}, SN~2010jl (discussed more below), SN2015da \citep{smith23da}, ASASSN-15ua \citep{dickinson24}, SN~2017hcc \citep{smith20}, and SN~2020ywx \citep{baerway25}.  SN~1998S also showed the blueshift in line profiles at late times more than a decade after explosion \citep{ms12,nicu22}. Recent {\it JWST}/MIRI spectroscopy of SN~2005ip revealed almost $\sim$0.1 $M_{\odot}$ of newly formed silicate dust in the CDS at $\sim$15 yr post-explosion \citep{shahbandeh25}. 

Six of these dusty SNe~IIn (SN~2006tf, SN~2010jl, SN~2015da, ASASSN-15ua, SN~2017hcc, and SN~2020ywx) were superluminous SNe~IIn (SLSNe~IIn) that resulted from especially strong CSM interaction with massive (5--25 $M_{\odot}$) CSM shells ejected in the decades before the explosion. Their extremely high luminosity results from efficient radiative cooling of the post-shock gas during the most intense interaction phase, so in hindsight, it is perhaps not surprising that their cooled post-shock layers may also be prodigious dust factories. It remains an open interesting question whether SLSNe~IIn are relatively more efficient dust producers than regular SNe~IIn. To answer this, a census of dust in samples of SNe~IIn and SLSNe~IIn is needed.

In this paper, we concentrate on SN~2010jl, a rare example of a relatively nearby well-observed SLSN~IIn, where observational clues were interpreted as post-shock dust formation. Located in the dwarf galaxy UGC~5189A at a distance of $\sim$49 Mpc \citep{smith11}, SN~2010jl was among the nearest SLSNe~IIn yet discovered, leading to considerable observational attention with high-quality optical/IR photometry and spectroscopy
\citep{andrews11,stoll11,smith11,smith12,zhang12,fox13,maeda13,gall14,fransson14,borish15,wf15,jencson16,tsvetkov16,sarangi18,chugai18,bevan20}. Its progenitor lived in a young region, implying that it had an initial mass above $\sim$30 $M_{\odot}$ \citep{smith11}, although the progenitor itself may have been very dust enshrouded \citep{fox17,dwek17,niu24}.

\begin{figure*}
\includegraphics[width=6.5in]{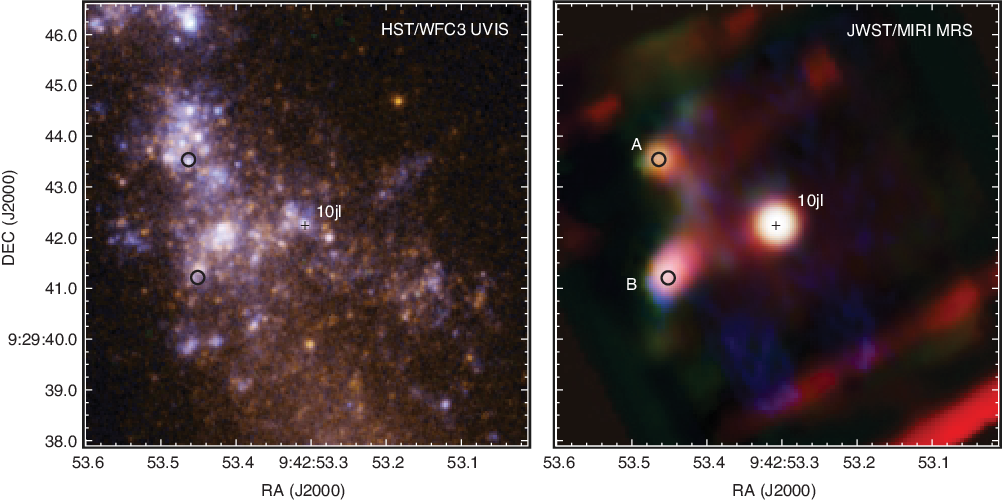}
\caption{Images of the field around SN~2010jl. The left panel shows a composite color {\it HST}/WFC3 UVIS image (blue = F336W, green = F336W+F814W, red = F814W) for reference.  The right panel displays the same field in a color image made from the {\it  JWST}/MIRI MRS data cube (blue = 5.8--10 $\mu$m, green = 10--15 $\mu$m, red = 15--25.6 $\mu$m). Examples of images made from the MIRI cube (program GO-1860; PI O. Fox) are also shown below in Figure~\ref{fig:astrobkginterp}. The F336W and F814W images from {\it HST}/WFC3 are averages of two epochs taken in 2015 (program GO-14149; PI A Filippenko) and 2016 (program GO-14668; PI A. Filippenko). The small black ``plus sign'' marks the position of SN~2010jl, while the two black open circles mark the positions of the two other bright MIR sources in the field (labeled A and B in the right panel), which are presumably dominated by massive star-forming regions in UGC~5189A.}
\label{fig:img}
\end{figure*}

\begin{figure*}
    \centering
    \includegraphics[width=0.95\textwidth]{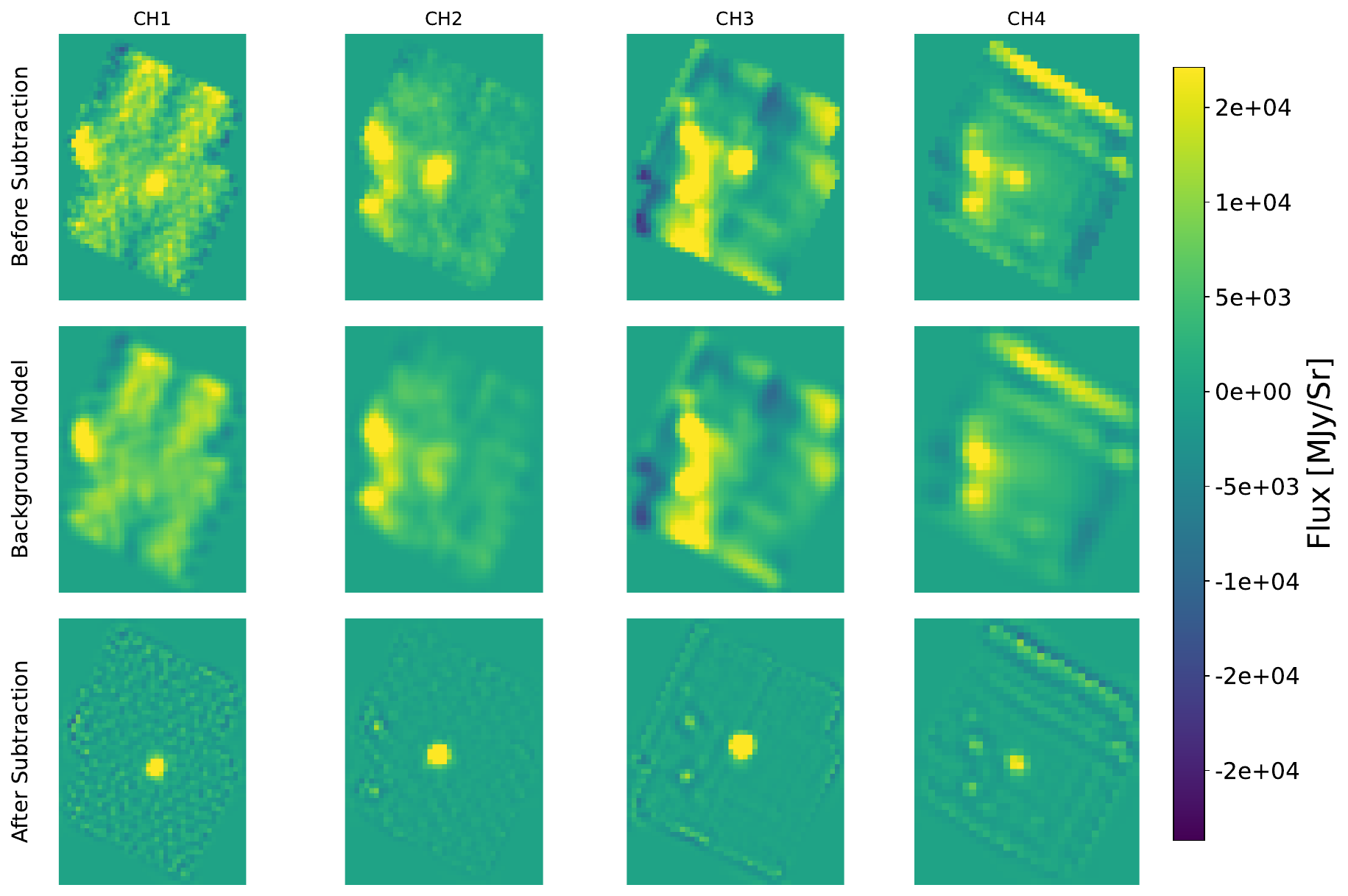}
    \caption{Input data, fitted background model, and background subtracted data from {\tt AstroBkgInterp} for four wavelength elements of the SN 2010jl MRS cube. The color bar corresponds to the flux in Jy of the background-subtracted panels. }
\label{fig:astrobkginterp}
\end{figure*}

\begin{figure*}\begin{center}
\includegraphics[width=0.99\textwidth]{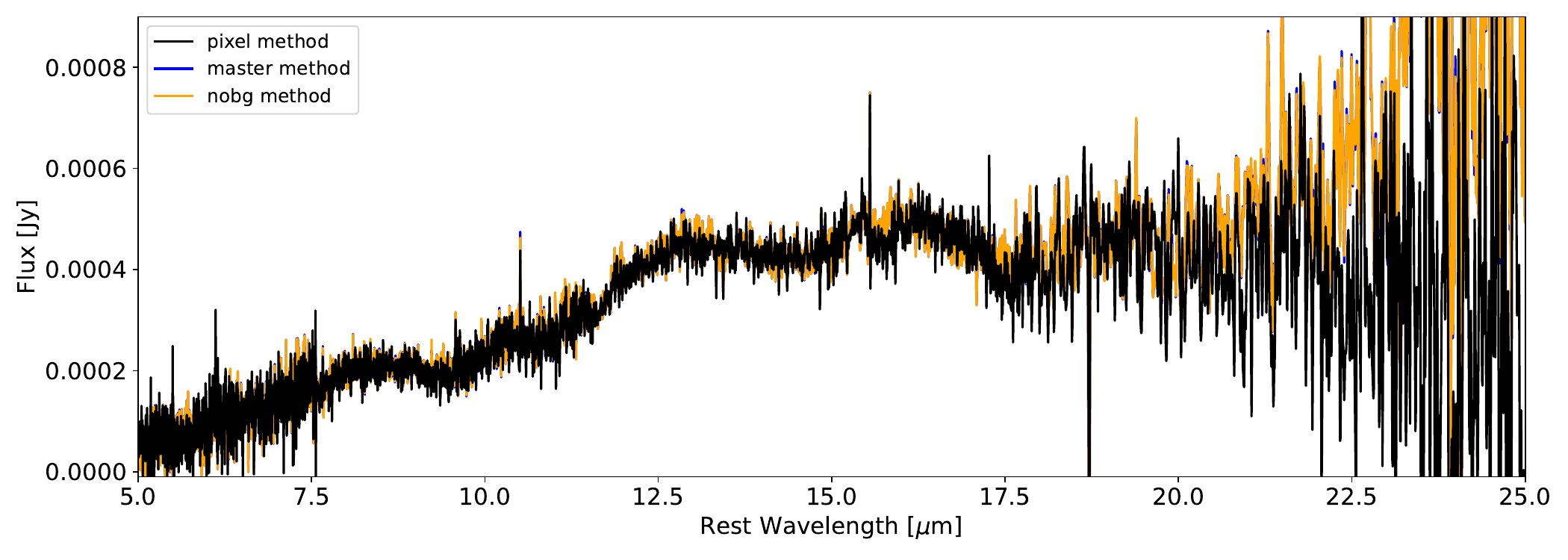}
\includegraphics[width=0.99\textwidth]{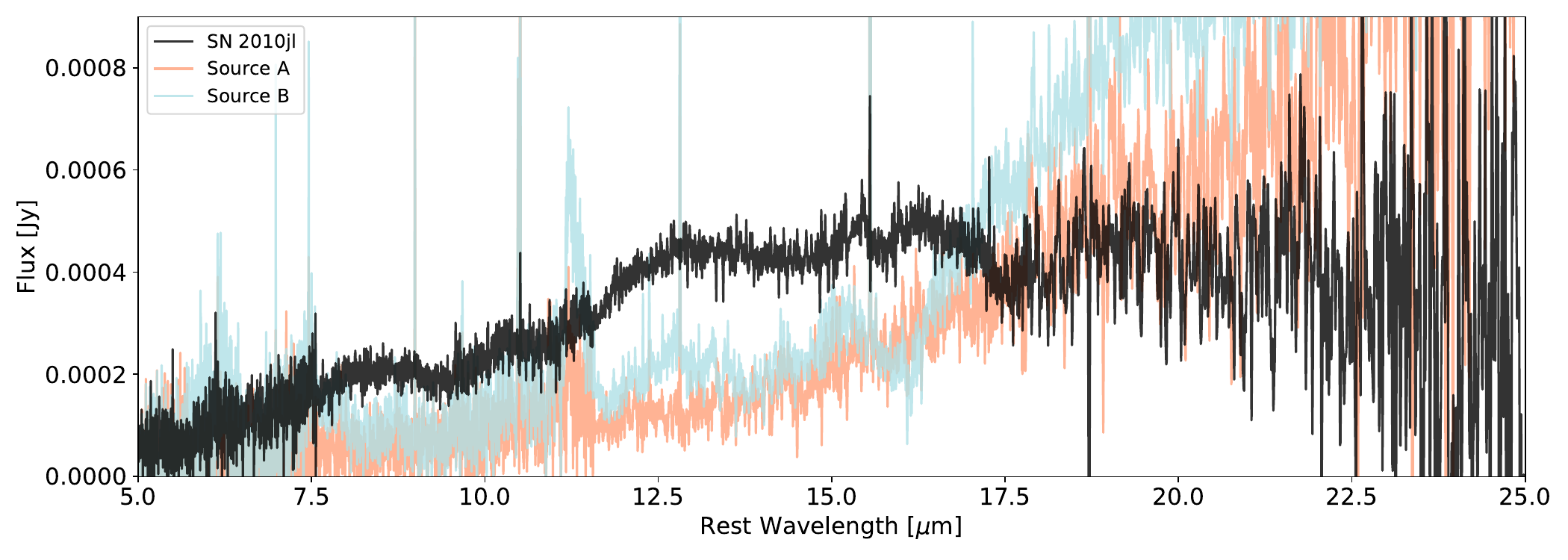}\end{center}
\caption{\textit{Top panel:} Examples of different reductions of the MIRI spectrum of SN~2010jl using three different background-subtraction methods (see text).  All three agree well at wavelengths shorter than 20 $\mu$m, but differences in method cause variations in the source flux in the range 20--25 $\mu$m. Our favored method of background subtraction (black spectrum) yields a lower 20--25 $\mu$m flux, and is used in the analysis below. 
\textit{Bottom panel:} A comparison between the spectrum of SN~2010jl (same as the black spectrum in the top panel) and spectra of the two H~{\sc ii} regions (sources A and B in Figure 1) in the MIRI MRS IFU.}
\label{fig:miri}
\end{figure*}

While pre-existing CSM dust probably contributed to the observed early-time IR
excess \citep{andrews11}, spectra in the year or so after the main light-curve peak revealed a
prominent blueshift in line profiles that strengthened over time \citep{smith12,gall14}.
The light curve of SN~2010jl also exhibited a pronounced inflection $\sim 300$ days after the explosion, although it was unclear if this faster decline rate was due to a drop in CSM interaction intensity, increased extinction, or both. Multiband photometry did show a strong increase in the IR excess emission as the optical faded after this inflection \citep{sarangi18,bevan20}.
The systematic blueshift of line profiles
and their wavelength dependence (more pronounced at shorter
wavelengths) led to the suggestion that, similar to SN~2006jc, SN~2010jl
experienced new dust formation in the post-shock region of the CDS
\citep{smith12}. Several additional studies confirmed these signs of
dust formation in the post-shock CDS and investigated the dust
properties, including additional late-time spectra, IR data, and
models \citep{gall14,sarangi18,chugai18,bevan20}. Dust may have been 
pre-existing in the CSM, and dust may also have
formed in the ejecta \citep{andrews11,bevan20}, but the wavelength dependence of the blueshift in IW profiles gave a clear indication of post-shock dust formation \citep{smith12,gall14}.  

\citet{fransson14} proposed a different picture where
the blueshifted profiles were caused by the acceleration of highly asymmetric pre-shock
CSM that had been ejected primarily toward Earth along our line of sight, and where the IW components were
caused by electron scattering of the narrow emission from that
accelerated CSM.  This explanation did not account for why the narrow
components from the unshocked CSM were not also blueshifted even though the
broader components had a blueshifted centroid, or why there was a
wavelength dependence to the asymmetry. Subsequent radiative-transfer
models \citep{dessart15} indicated that
accelerated pre-shock CSM could not explain the observed blueshift in
line profiles, nor their behavior with time. \citet{dessart15}
found that while the early symmetric line profiles were caused
by electron scattering of narrow CSM emission, the broader blueshifted
profiles at later times probably came from the post-shock CDS. These
models revealed a blueshifted emission bump that could arise even
without dust, at least initially, because the
photosphere in the CSM interaction region could block the redshifted
side of the CDS, as noted earlier by \citet{smith12}. However, again,
this mechanism would not account for the observed wavelength
dependence of the blueshift \citep{smith12,gall14}, or its persistence to late times, and the behavior of the observed blueshift over time did not resemble the development of the blueshifted bump in the models. The continuum luminosity dropped around day 300, as noted above, but the
blueshift persisted even in optical spectra beyond day 1000
\citep{fransson14}. This would seem to clearly rule out high
continuum optical depths and electron scattering of accelerated CSM as
the explanation for the persistent blueshift, and instead, favors dust formation in
the post-shock shell of SN~2010jl \citep{smith12,gall14,sarangi18,bevan20}.

While the blueshifted line profiles provided a strong indication of new dust forming within the SN itself, and they help diagnose its location, it remains difficult to estimate the mass of dust from this velocity-dependent extinction alone, even with line-profile modeling \citep{bevan20}. A combination of spectra and IR emission is needed. Early visual-wavelength and near-IR data showed evidence for a rapid onset of dust formation \citep{smith12,gall14}, and line-profile data before day 1000 showed that the observationally inferred dust mass continued to grow with time, reaching $2.5 \times 10^{-3}$ $M_{\odot}$ by day 868 \citep{gall14}. Similarly, by modeling line-profile shapes, \citet{bevan20} estimated a dust mass of 0.005--0.01 $M_{\odot}$ by day 1400.  Here we revisit the visual-wavelength and IR emission from SN~2010jl a decade later, to see if the dust mass has continued to grow as was apparently the case for SN~1987A and SN~2005ip. This study includes late-time optical spectra from Keck and the MMT, and MIR spectra obtained with \jwst. We present the new spectra in Section~\ref{sec:obs} and our analysis of the spectra in Section~\ref{sec:analysis}. The results are discussed in the context of dust formation by interacting SNe in Section~\ref{sec:discussion}.

\begin{deluxetable}{ l l l l l l}
\setlength{\tabcolsep}{0.8pt}
\small
\caption{{\it JWST}/MRS Spectroscopy of \jl \label{tab:tab1}}
\tablehead{
\colhead{Channel} & \colhead{FOV} & \colhead{Band} & \colhead{Wavelength} & \colhead{Resolving Power} & \colhead{Exp.} \\
\colhead{}     &    \colhead{(\arcsec $\times$ \arcsec)} & \colhead{} & \colhead{(\micron)} & \colhead{($\lambda/\Delta\lambda$)} & \colhead{(s)}
    }
\startdata
\hline
\multirow{3}{*}{1} & \multirow{3}{*}{3.2$\times$3.7} & Short  & 4.90–5.74 & 3,320–3,710 & 1337\\
                   &                                  & Medium & 5.66–6.63 & 3,190–3,750 & 1337\\
                   &                                  & Long   & 6.53–7.65 & 3,100–3,610 & 1337\\
\hline
\multirow{3}{*}{2} & \multirow{3}{*}{4.0$\times$4.8} & Short  & 7.51–8.77 & 2,990–3,110 & 1337\\
                   &                                  & Medium & 8.67–10.13 & 2,750–3,170 & 1337\\
                   &                                  & Long   & 10.02–11.70 & 2,860–3,300  & 1337\\
\hline
\multirow{3}{*}{3} & \multirow{3}{*}{5.2$\times$6.2} & Short  & 11.55–13.47 & 2,530–2,880 & 1337\\
                   &                                  & Medium & 13.34–15.57 & 1,790–2,640 & 1337\\
                   &                                  & Long   & 15.41–17.98 & 1,980–2,790  & 1337\\
\hline
\multirow{3}{*}{4} & \multirow{3}{*}{6.6$\times$7.7} & Short  & 17.70–20.95 & 1,460–1,930 & 1337\\
                   &                                  & Medium & 20.69–24.48 & 1,680–1,770 & 1337\\
                   &                                  & Long   & 24.19–27.90 & 1,630–1,330  & 1337\\
\hline
\enddata
\end{deluxetable}

\section{OBSERVATIONS}\label{sec:obs}

\subsection{\jwst/MIRI MRS}\label{sec:data_jwst}

{\it JWST}/MIRI's Medium Resolution Spectrometer (MRS; \citealt{wells15, bushouse22,argyriou23}) integral field unit (IFU) observed \jl\ on 2023 April 19 (JD 2,456,401), 4,550 days post-discovery (UTC dates are used throughout this paper). The data consist of spectra covering the wavelength range 4.9--27.9~\micron, aimed at measuring the dust emission. Table \ref{tab:tab1} summarizes the observation parameters. These observations are part of PID 1860 (PI O. Fox), which probes the dust properties around SNe~IIn. 

For these observations, it is necessary to subtract both the global thermal background and local galaxy background. The global thermal background is roughly 10--40 times larger than the SN flux. We obtained dedicated offset background observations to provide a clean estimate of the telescope's thermal background. Even so, small variability in the background across the field of view (FOV) can have a potentially large impact on our source spectrum \citep{shahbandeh24}. This is particularly true at the longest wavelengths, where the background tends to be the largest, and the signal-to-noise ratio (S/N) from the source tends to be the weakest owing to a combination of lower signal and the known reduced count rate.\footnote{\url{https://www.stsci.edu/contents/news/jwst/2023/miri-mrs-reduced-count-rate-update}}  The removal of the global thermal background does not include any estimate of the local background from SN~2010jl's host galaxy. While the individual sources in the {\it HST}/WFC3 image are not resolved in the MRS data, they do create faint, extended emission, often referred to as ``confusion-limited.'' 

For this paper, we explored several different techniques to remove the background. We first addressed the global thermal background by reducing the data three different ways: (1) with no global background subtractions (``no bg''), with an average background applied using our dedicated offsets (``master method''), and a spaxel-by-spaxel background subtraction using our dedicated offsets (``pixel method''). Level-1 data were initially downloaded from MAST.\footnote{\dataset[DOI:10.17909/b3f9-n162]{\doi{10.17909/b3f9-n162}}} All data were processed with {\it JWST} calibration pipeline version 1.11.1 \citep{bushouse22}, using CRDS pmap 1183. We follow the standard MRS pipeline procedure \citep{bushouse24}. The background-subtraction step can be applied automatically in the {\it JWST} pipeline.\footnote{https://jwst-pipeline.readthedocs.io/en/stable/jwst/user\_documentation/ background\_subtraction\_methods/main.html\#master-background-subtraction} 

At the local background level, we experimented with removing the remaining host-galaxy background, including implementing a background annulus during the extraction, averaging an array of apertures around the FOV, and performing a two-dimensional (2D) interpolation and estimate of the local background. We ultimately found the 2D interpolation to be most robust. This technique is similar to the background-estimation step used in HYPER \citep{traficante15}. The custom 2D interpolation code \citep{nickson}
used here is a highly parameterized Python module that creates a model of the entire background for a source of interest in a complex region. The source is masked by interpolating over an aperture using the surrounding sky. The background is modeled by iteratively using Bayesian inference to fit a 2D polynomial to smaller substamps of the masked image.   The uncertainty introduced by this routine is folded into the flux error.  We selected a substamp size of 7$\times$7 pixels, fit a second-order polynomial to each, and set the step-size parameter, {\tt cube\_resolution}, to ``high." Figure~\ref{fig:astrobkginterp} shows the input data, fitted background model, and background subtracted image for four example wavelength elements of the S3D data cube.

The SN spectrum was extracted from the resulting data cube using the \jwst~pipeline \texttt{extract\_1d} step. No additional background subtraction (such as an annulus) was performed, as the 2D interpolation was considered to have adequately removed the local background at the position of the source. For point sources, the pipeline uses a circular extraction aperture, which varies in radius with wavelength to take into account the changing resolution. 

Figure~\ref{fig:miri} (top panel) shows the resulting spectra for each of the three approaches to removing the thermal background (using the same local background 2D interpolation subtraction method). Each of the methods yields similar results at wavelengths below 20 \micron, which is expected as the thermal background is less significant there. At longer wavelengths, the ``pixel'' method (black line) yields a slightly lower flux. This  result is most likely due to the fact that the ``pixel'' method properly accounts for spaxel-by-spaxel variations across the FOV. It also provides a robust lower limit to our models. We therefore take this extraction as our final extraction throughout the rest of the paper.

We note that the spectrum shows a pronounced dip in flux at 17--17.5~\micron, which has important implications in our models discussed below. This dip also corresponds to the start of Channel~4. To check if this 17 $\mu$m dip is an instrumental effect, we also extract spectra for two nearby sources in the FOV (labled as A and B in Figure~\ref{fig:img}) using the same method as for \jl. These two sources are most likely dominated by warm dust surrounding H~{\sc ii} regions. The purpose of the comparison is to look for any overall trends in the spectra that may indicate a systematic instrumental source of error. Figure~\ref{fig:miri} (bottom panel) compares \jl\ to these two sources. Sources A and B do not show a clear dip at 17 $\mu$m and their flux rises smoothly to longer wavelengths, so this may be a real feature in SN~2010jl's MIR spectrum.  

Nevertheless, as detailed above, Channel 4 has challenges related to background subtraction. Furthermore, it is known to experience a progressive reduction in the count rate with time. As a result, we regard the flux beyond 17 $\mu$m with some caution. We place less emphasis on the detailed features at these wavelengths in our modeling below, although the overall flux level represented by the binned flux is still meaningful. In hindsight, it would have been beneficial to have also obtained imaging at 20--25 $\mu$m to better anchor the dust continuum level, since it is so important for the resulting derived dust mass.

\begin{figure*}\begin{center}
\includegraphics[width=6.5in]{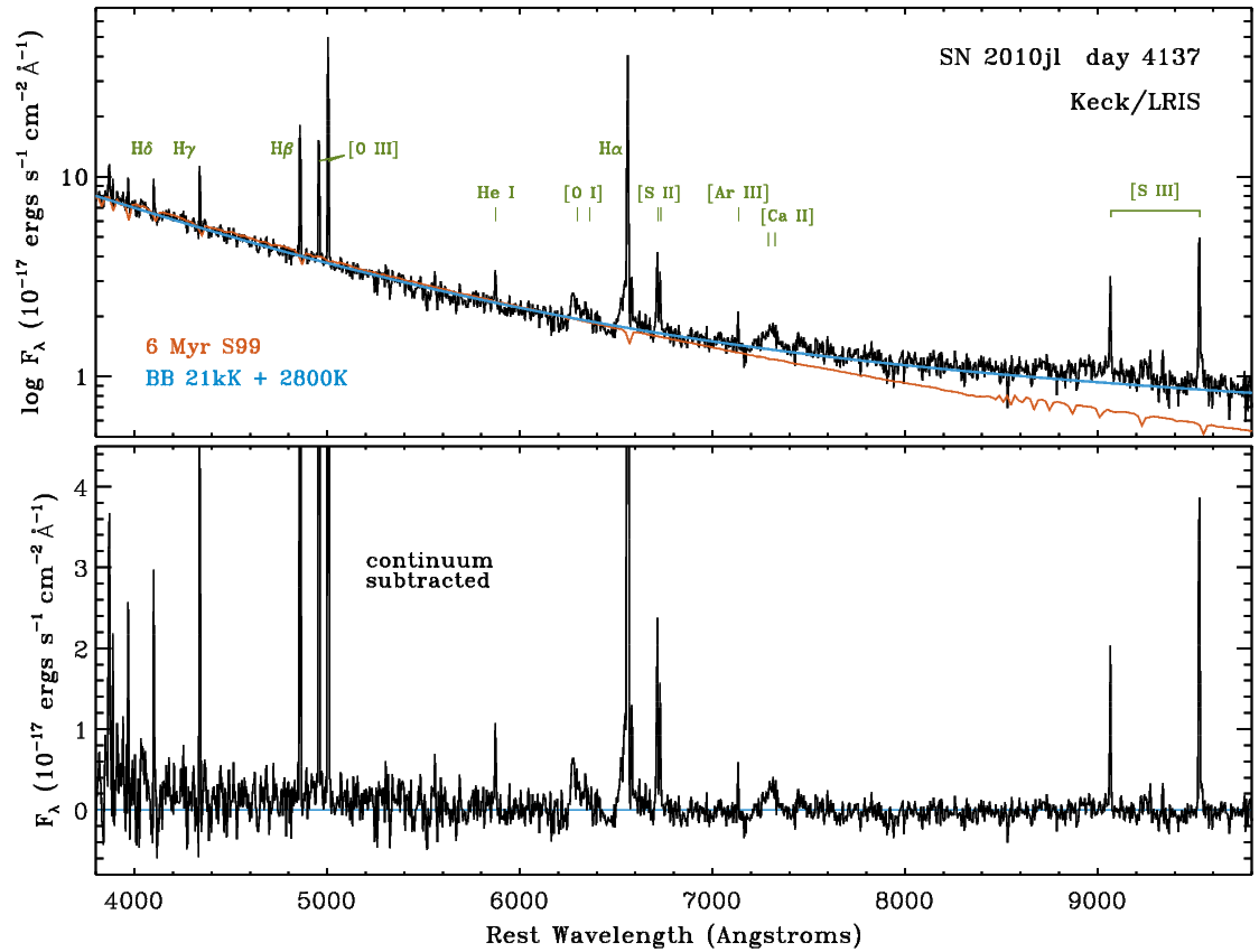}\end{center}
\caption{Late-time (day 4137) Keck/LRIS spectrum of the source at the position of \jl. The top panel shows the original flux-calibrated spectrum on a log flux scale.  Several emission lines (mostly unresolved narrow emission from a coincident H~{\sc ii} region) are identified in green. For comparison, the red spectrum is the same Starburst99 (S99) model \citep{starburst99} for a 6 Myr cluster that was matched to the {\it HST} progenitor photometry by \citet{smith11}. This model matches the observed blue continuum flux very well but significantly underestimates the near-IR flux longward of 7000 \AA. Also shown for comparison, the blue curve is the composite of two blackbodies at 21,000 and 2,800 K. This is artificial, but gives a better approximation of the overall continuum distribution, indicating some additional cool source that is not seen in a model of a 6 Myr cluster, and was not seen in the progenitor photometry.  The bottom panel displays the same spectrum as the top panel, except that the smooth continuum (blue curve in the top panel) has been subtracted.  Here, the only features to show unambiguous shock-broadened SN emission are [O~{\sc i}] $\lambda\lambda$6300, 6364, H$\alpha$, and [Ca~{\sc ii}] $\lambda\lambda$7291, 7325. Besides these broadened components, most of the late-time flux at the position of SN~2010jl is due to an associated star cluster and H~{\sc ii} region.}
\label{fig:spec}
\end{figure*}

\begin{figure*}\begin{center}
\includegraphics[width=6.0in]{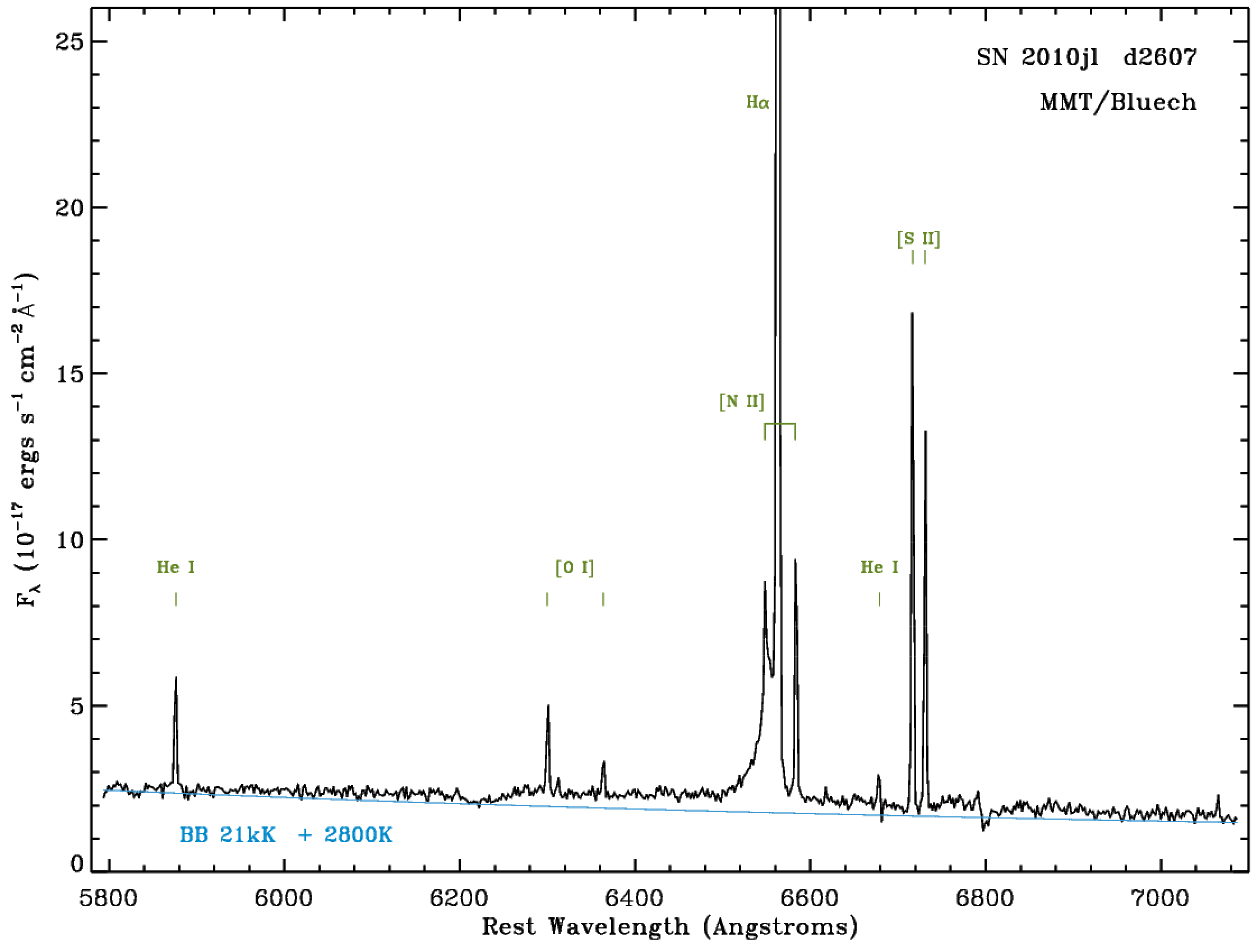}\end{center}
\caption{Same as Fig.~\ref{fig:spec}, but for the MMT/Bluechannel spectrum on day 2607. The blue line is not a fit to the continuum here, but is identical to the blue curve representing the two-blackbody curve in Fig.~\ref{fig:spec}. The somewhat brighter continuum here may be real (i.e., some late-time SN continuum emission), but is more likely attributable to a difference in seeing or slit position and therefore slit losses.  The narrow H~{\sc ii} region lines appear stronger here because of the 4$\times$ higher spectral resolution than the LRIS spectrum. }
\label{fig:specMMT}
\end{figure*}

\subsection{Optical Spectra}

Several previous studies have presented the visual-wavelength spectral evolution of \jl\ during its main luminosity peak and decline \citep{smith11,smith12,zhang12,fransson14,jencson16}. After a few years, however, SN~2010jl dropped significantly in brightness. There have been no studies of the very late-time emission at 6 or more years after explosion. The latest published visual-wavelength spectra were obtained on roughly day 1287 \citep{bevan20} and day 1909 \citep{nicu22}.

We obtained a deep spectrum of the source at the position of SN~2010jl on 2022 March 3, which is $\sim 4137$ days after discovery.  This is the latest visual-wavelength spectrum available to us, and the closest to being contemporaneous with our {\it JWST} data. Fortunately, the spectral evolution of SNe~IIn at very late times is expected to be very slow \citep{smith17}, showing only subtle changes over a year. This spectrum was obtained with the 10\,m Keck-I telescope using the Low-Resolution Imaging Spectrometer \citep[LRIS;][]{oke95} plus atmospheric dispersion corrector, and had an exposure time of 1800~s using a 1{\arcsec}-wide slit. The spectral resolving power is $R \equiv \lambda/\Delta \lambda \approx 1100$, corresponding to a resolution of 273 km s$^{-1}$. Data reduction used the LPipe pipeline \citep{perley19}. We applied a redshift correction of $z=0.0107$ and a reddening correction of $E(B-V) = 0.027$~mag \citep{smith11}. This LRIS spectrum is plotted in Figure~\ref{fig:spec}.  The top panel is the flux-calibrated spectrum, and the bottom panel shows it with the smooth continuum subtracted. 

We also obtained a late-time spectrum with the Bluechannel spectrograph on the MMT Observatory on 2017 Dec. 23, which is 
$\sim 2607$ days after discovery. This MMT spectrum is not as late in time as the Keck/LRIS spectrum and covers a smaller wavelength range, but it is useful because it has a comparable continuum S/N 
and higher resolution than the Keck/LRIS spectrum, thus providing a better measure of the shock-broadened component of H$\alpha$. This observation used a 1{\arcsec}-wide slit at the parallactic angle \citep{filippenko82}, and the 1200 line mm$^{-1}$ grating, providing a spectral resolving power of $R \approx 4400$, which corresponds to the unresolved $\sim$70 km s$^{-1}$ width of the narrow H~{\sc ii} region emission lines in the spectrum. We used standard data-reduction procedures for single-order long-slit spectra. The MMT/Bluechannel spectrum is shown in Figure~\ref{fig:specMMT}.

In both cases, the 1{\arcsec} slit  (note that 1{\arcsec} is about 240 pc at the distance of SN~2010jl) used in these optical spectra will include nearby nebular H~{\sc ii} region emission, which is bright and pervasive in the surroundings of SN~2010jl, and will also include the blue source located only 0$\farcs$06 ($\sim$15 pc) away from the position of SN~2010jl, plus some of the surrounding stellar population. This source was resolved in pre- and post-explosion {\it HST} images and is thought to be a young star cluster \citep{smith11,fox17}, as further discussed in Section~\ref{sec:analysis}.

\begin{figure}\begin{center}
\includegraphics[width=3.0in]{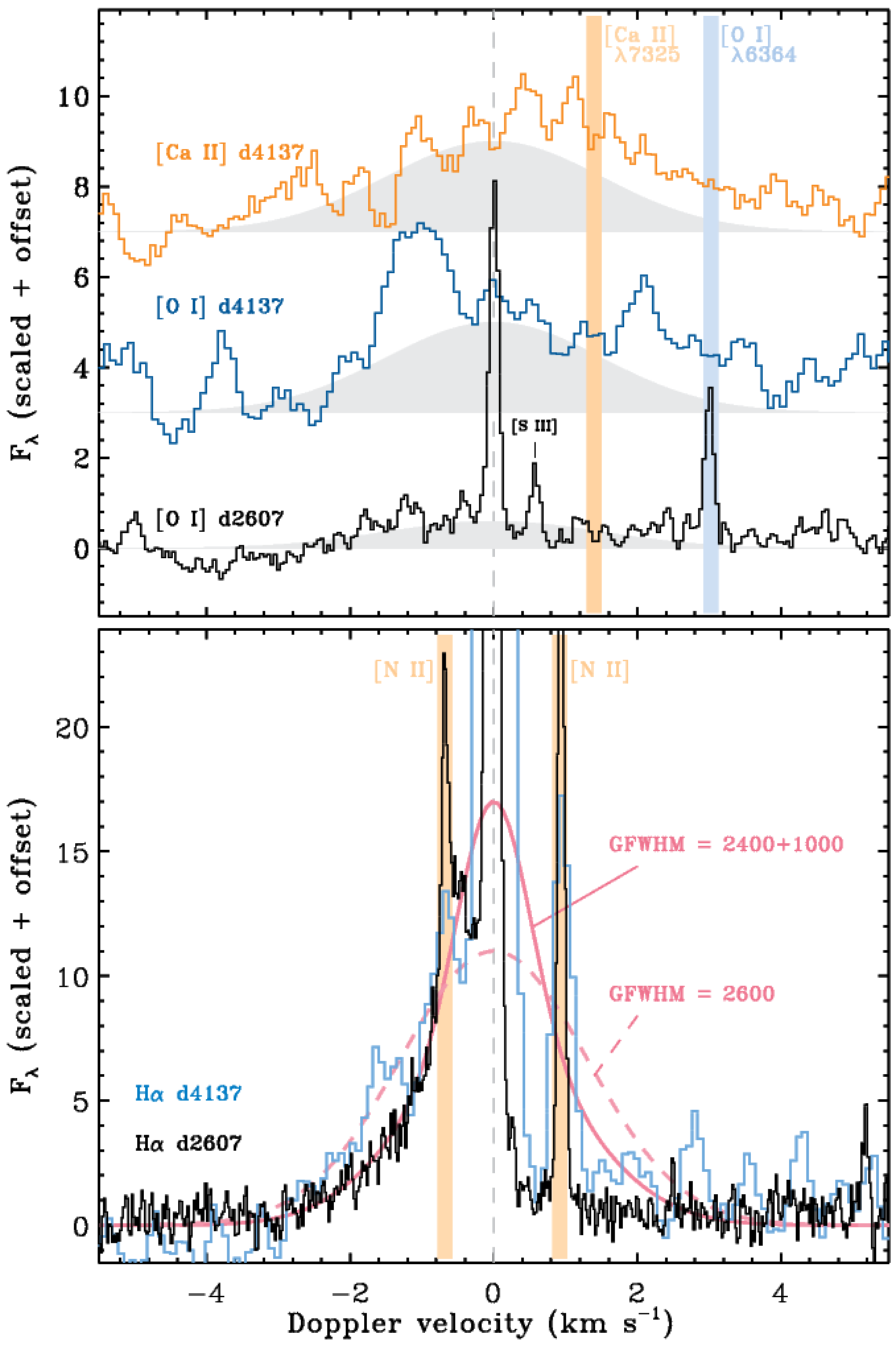}\end{center}
\caption{Line profiles for shock-broadened emission in the optical spectra. The top panel shows [O~{\sc i}] and [Ca~{\sc ii}] in the day 4137 Keck spectrum in blue and orange, respectively, and the [O~{\sc i}] profile in the day 2607 MMT spectrum in black.  Note that both [O~{\sc i}] and [Ca~{\sc ii}] are doublets; in both cases we set $v=0$ to be the shorter-wavelength line in each doublet.  The gray shaded regions show Gaussians with FWHM = 2600 km s$^{-1}$ for comparison.  The bottom panel displays the H$\alpha$ profile in the MMT (black) and Keck (blue) spectra.  The adjacent narrow [N~{\sc ii}] lines are marked in orange.  The pink curves show representative symmetric Gaussian curves with FWHM values noted in the panel; the dashed curve is for comparison with the blue wing of H$\alpha$ in the Keck spectrum, and the solid curve matches the blue wing of H$\alpha$ in the MMT spectrum.  In both observed spectra, the redshifted emission from these broad components is obviously missing compared to a symmetric profile.}
\label{fig:vel}
\end{figure}

\section{ANALYSIS}\label{sec:analysis}

\subsection{Late-Time Optical Spectra}

Figures~\ref{fig:spec} and \ref{fig:specMMT} show our late-time visual-wavelength spectra obtained 7--12 yr after discovery. These are the latest optical spectra of SN~2010jl, and among the latest spectra of any SLSN~IIn (a spectrum of the SLSN IIn event SN~2015da obtained  8.4 yr after the explosion was presented by \citealt{smith23da}).  The spectra show clear signatures of ongoing late-time CSM interaction in SN~2010jl, although these signatures are faint, and the spectra are dominated by flux from the surrounding environment.

The day 4137 Keck/LRIS spectrum in Figure~\ref{fig:spec} is dominated by narrow emission lines and a smooth blue continuum. The narrow lines are unresolved, and the spectrum is typical of emission from low-metallicity H~{\sc ii} regions, with strong Balmer lines, [O~{\sc iii}], [S~{\sc ii}], and [S~{\sc iii}], and fainter narrow He~{\sc i}, [N~{\sc ii}], [Ar~{\sc iii}], and [O~{\sc i}].  

The blue continuum is dominated by flux from an underlying young star cluster that is almost coincident with SN~2010jl \citep{fox17}. The blue source in pre-explosion {\it HST} images that is located only 0$\farcs$06 (15 pc) away in projection \citep{smith11} is easily included in the ground-based slit. The thin red continuum spectrum plotted in the top panel of Figure~\ref{fig:spec} is a Starburst99 model \citep{starburst99} of the integrated flux from a 6 Myr cluster.  This is not a fit to these data, but is instead the same model that was compared to the pre-explosion photometry of the blue source at SN~2010jl's approximate position \citep{smith11}.  This Starburst99 model provides an excellent match to the blue continuum  in the Keck spectrum, but it underestimates the observed continuum  at near-IR wavelengths beyond 7000 \AA. Alternatively, we can find a better match to the overall continuum shape using a combination of two Planck functions with temperatures of 21,000 and 2,800 K. This is shown with the blue curve in the top panel of Figure~\ref{fig:spec}.  While this composite blackbody curve matches the continuum shape better than the cluster model, it is admittedly artificial, and we do not wish to overinterpret the temperatures.  The 21,000 K source\footnote{Note that the exact temperature of this soure, and similarly, the age of the cluster in the Starburst99 model, are dependent on the adopted $E(B-V)$ value for the reddening correction applied to the spectrum.} is simply indicative of the Rayleigh-Jeans tail from the combination of hot stars in a young cluster, while the 2,800 K blackbody captures extra red emission from some source(s) in the aperture. The progenitor photometry \citep{smith11} did not show this red excess, but the progenitor photometry was measured for a blue point source in {\it HST} images with a resolution of $\sim$0$\farcs$1. These same images show several nearby stellar sources that are brighter in the F814W image than at blue/near-ultraviolet (UV) wavelengths, and several of these sources would be included in the 1{\arcsec} slit aperture used for the ground-based spectrum. Thus, part of the $\sim$2,800 K extra red continuum flux seen in the Keck spectrum in Figure~\ref{fig:spec} may be attributable to some red supergiants in neighboring stellar associations near SN~2010jl; however, we cannot rule out the possibility that some of the long-wavelength emission might also be due to hot dust associated with the SN, since a strong dust excess is detected at MIR wavelengths (see below). At earlier epochs before day 900, \citet{gall14} estimated that hot dust contributed significant red continuum, so this contribution may have persisted if hot dust is continually present.  Hot dust can be seen even after more than a decade if an advancing shock front continually heats it.

The bottom panel of Figure~\ref{fig:spec} shows the residual after subtracting a fit to the continuum (blue curve in the top panel).  The narrow H~{\sc ii} region lines remain, of course, but here it is easier to see the only features in the spectrum that are clearly attributable to the late SN emission: this spectrum reveals shock-broadened emission from [O~{\sc i}] $\lambda\lambda$6300, 6364, H$\alpha$, and [Ca~{\sc ii}] $\lambda\lambda$7291, 7325. These components have widths of 2000--3000 km s$^{-1}$ (usually referred to as ``intermediate-width" or IW components), much broader than the $\sim$300 km s$^{-1}$ width of the unresolved H~{\sc ii} region lines, and likely arise from post-shock gas in the aging SN (or, alternatively, in the young SN remnant).  The broad line profiles are discussed more below.  Interestingly, the Keck/LRIS spectrum does not detect any emission from the blend of O~{\sc i} $\lambda$8446 and the Ca~{\sc ii} near-IR triplet, which was seen in earlier spectra of SN~2010jl \citep{smith12,fransson14}, although these features became very weak by day 909 \citep{jencson16}.  This broad blend was also seen in emission in late-time spectra of some other SLSNe~IIn such as SN~2006tf \citep{smith08tf} and SN~2015da \citep{smith23da}.

The MMT/Bluechannel spectrum is shown in Figure~\ref{fig:specMMT}.  This was obtained in 2017, about 4 yr before the Keck spectrum.  It covers a smaller wavelength range and is not as close in time to our {\it JWST} spectrum, but it has better spectral resolution and higher S/N than the Keck spectrum, and is especially useful for investigating the H$\alpha$ line profile, as discussed below.  The blue curve in Figure~\ref{fig:specMMT} is the same two-blackbody curve that matched the continuum shape in the Keck spectrum.  Although the wavelength range here is smaller, the MMT spectrum suggests that the underlying continuum changed very little over the 4 yr between the MMT and Keck spectra, consistent with the idea that this continuum flux is dominated by the nearby young star cluster and not SN continuum emission.  Comparing the two spectra does reveal some interesting changes in SN emission; namely, the [O~{\sc i}] line is much fainter in the MMT spectrum, and the shock-broadened component of H$\alpha$ is narrower at earlier times in the MMT spectrum, although both spectra show strongly asymmetric blueshifted H$\alpha$, discussed more below.  The MMT spectrum does not extend to sufficiently long wavelengths to see [Ca~{\sc ii}] $\lambda\lambda$7291, 7325.

The behavior of [O~{\sc i}] $\lambda\lambda$6300, 6364 is noteworthy.  This line was not detected in earlier spectra up to 1200 days after discovery \citep{smith11,smith12,fransson14,jencson16}.  The emission is extremely faint or not detected in our MMT spectrum on day 2607, but then it is clearly detected as shock-broadened emission (about half the flux of H$\alpha$) in the day 4137 Keck spectrum.  This strengthening is interesting, since the general absence of evidence for [O~{\sc i}] in late-time spectra of SNe~IIn might erroneously be taken to indicate that they are not core-collapse explosions, or alternatively, weak [O~{\sc i}] emission might be used to derive a low progenitor initial mass.  Normal SNe II without interaction do exhibit [O~{\sc i}] in their nebular phases.  Saganian wisdom  aside, there are two good reasons why such arguments based on missing [O~{\sc i}] would be flawed.  One reason is that ongoing late-time interaction can obviously mask or overwhelm the nebular emission, and backwarming by shock radiation may even alter the underlying nebular emission, ionizing oxygen to O$^+$  \citep{smith14ip}.  But a second reason, seen here, is that one might simply need to wait long enough for the line to appear as the CSM interaction emission fades at very late times.  

The line profiles of the shock-broadened emission in our late-time spectra are shown in detail in Figure~\ref{fig:vel}.  Interpreting the line profiles of [O~{\sc i}] and [Ca~{\sc ii}] (shown in the top panel of Fig.~\ref{fig:vel}) is complicated because these lines are relatively weak and the spectra are noisy, and more importantly, because both features are doublets.  The lines of the [O~{\sc i}] $\lambda\lambda$6300, 6364 doublet are separated by $\sim 3,000$ km s$^{-1}$, and [Ca~{\sc ii}] $\lambda\lambda$7291,7325 by roughly 1,400 km s$^{-1}$.  (For both features, we set zero velocity by the shorter-wavelength line in each pair; vertical blue and orange bars in Fig.~\ref{fig:vel} mark the corresponding velocities of the companion line centers in each doublet.)  Despite these limitations, it is clear that there is some shock-broadened emission detected in both lines in the day 4137 Keck spectrum, while the data permit some very faint broadened [O~{\sc i}] emission in the MMT spectrum.  Symmetric Gaussians with full width at half-maximum intensity (FWHM) = 2600 km s$^{-1}$ are shown for comparison in shaded gray in the top panel of Fig.~\ref{fig:vel}, although the velocity width may be less than this because both features are a blend of two lines.  It is difficult to see if the lines have a blueshifted asymmetry for this same reason, athough [O~{\sc i}] does show an especially strong blue emission bump at around $-$1000 km s$^{-1}$.  Spectra that cover [O~{\sc i}] $\lambda\lambda$6300, 6364 and [Ca~{\sc ii}] $\lambda\lambda$7291, 7325 with higher S/N and greater spectral resolution are needed.  Although the MMT spectrum provides this for [O~{\sc i}], this line did not strengthen in its emission until after that spectrum was obtained.

On the other hand, the broadened component of H$\alpha$ clearly displays the blueshifted asymmetric line profile that is often seen in SNe IIn, shown here to persist until very late times more than a decade after explosion.  This blueshifted H$\alpha$ in SN~2010jl first started to appear as a subtle asymmetry around 60--70 days after discovery \citep{smith11}, and only became more pronounced over time \citep{smith11,smith12,gall14,bevan20,nicu22}, persisting until the present epoch when essentially no emission is seen from the redshifted side of the line.  The bottom panel of Figure~\ref{fig:vel} shows H$\alpha$ in detail.  Ignoring the narrow [N~{\sc ii}] $\lambda\lambda$6548, 6583 lines from the nearby H~{\sc ii} region emission in the slit aperture (and by extension, the very strong, narrow H$\alpha$ emission), the broader H$\alpha$ component shows a smooth blue wing that rises to line center, but then appears to be chopped off at zero velocity, with no detectable broad emission on the red side of the line.  If we attempt to fit only the shape of the blue wing, the day 4137 Keck profile roughly matches a Gaussian with FWHM = 2600 km s$^{-1}$ (dashed pink curve).  The MMT spectrum, where the H$\alpha$ profile is seen with better S/N and higher resolution, a single Gaussian is not able to match the blue wing; instead, a composite of two Gaussians with FWHM values of 2400 and 1000 km s$^{-1}$ (roughly with equal strength; shown with a solid pink curve) give a much better match to the shape of the blue wing.  These pink curves in the bottom panel of Figure~\ref{fig:vel} illustrate the severe deficit of flux on the red wing of H$\alpha$.

It is perhaps somewhat surprising that the IW component of H$\alpha$ appears to get broader in the 4 yr between the MMT spectrum and the Keck spectrum, with the implied FWHM increasing.  This would seem to suggest that the speed of the forward shock, coincident with the CDS,  has actually accelerated during this time.  A similar acceleration of the forward shock, traced by an increasing width of the IW component, was also reported recently in late-time spectra of SN~2015da \citep{smith23da}, another SLSN IIn with even stronger late-time interaction than SN~2010jl.

In the context of trying to decipher the location of dust that blocks the redshifted emission, which will be discussed more below, the detailed profile shape around zero velocity is of interest.  The IW emission component arises from gas in the post-shock region (the CDS).  The red wing of this IW component could be absorbed, at least in principle, by dust located either in the post-shock CDS or in the freely expanding SN ejecta inside the reverse shock.  However, several authors have noted that if the flux of the IW component is significantly attenuated at zero velocity or even to modest blueshifted velocities (shifting the peak to blueshifted velocities), then the dust must be located in the post-shock region, because dust in the inner parts of the SN ejecta cannot block post-shock gas exterior to it moving in the plane of the sky \citep{smith08jc,smith08tf,smith20,gall14,ss22}.  A close examination of the day 2607 MMT spectrum in the window between the narrow [N~{\sc ii}] $\lambda$6548 and narrow H$\alpha$ emission shows a peak at roughly $-$500 km s$^{-1}$, making it appear as though the narrow H$\alpha$ component sits atop a steep slope as the underlying broader H$\alpha$ component drops toward zero flux on the red side of the line.  Given the data quality, this might seem dubious if it were not for the fact that the IW component of H$\alpha$ has shown a similar profile with a blueshifted peak at roughly $-$500 km s$^{-1}$ ever since 200--300 days after explosion \citep{smith12,gall14}.  In other words, while the flux has dropped over time, the shape of the IW component of H$\alpha$ has remained relatively unchanged for the past decade. As we discuss in the next section, this confirms that the dust blocking the red side of the line is in the post-shock CDS, and not in the SN ejecta.

While the IW component of H$\alpha$ has remained the same for over a decade, the broad wings of H$\alpha$ have not.  In these late-time MMT and Keck spectra, there is no detected blueshifted emission faster than about $-$3000 km s$^{-1}$, and there is no redshifted emission at all.  In earlier spectra up to a few years after the explosion, H$\alpha$ exhibited broad wings to more than $\pm$6,000 km s$^{-1}$ \citep{smith12,gall14,fransson14,jencson16}, with roughly symmetric profiles at the highest velocities.  If the broad wings were tracing the fastest SN ejecta, then their absence now suggests that either the inner SN ejecta have cooled and faded, or that the faster SN ejecta have reached the reverse shock and have been decelerated.  Alternatively, if the broad wings at earlier times were due in part to electron scattering, then their absence now would be consistent with the disappearance of any significant electron-scattering opacity.  In any case, only the IW components emitted by the CDS remain detectable in the spectrum, which is common in very late spectra of SNe IIn \citep{smith17ip}.

\begin{figure*}
    \centering
    \includegraphics[width=0.45\textwidth]{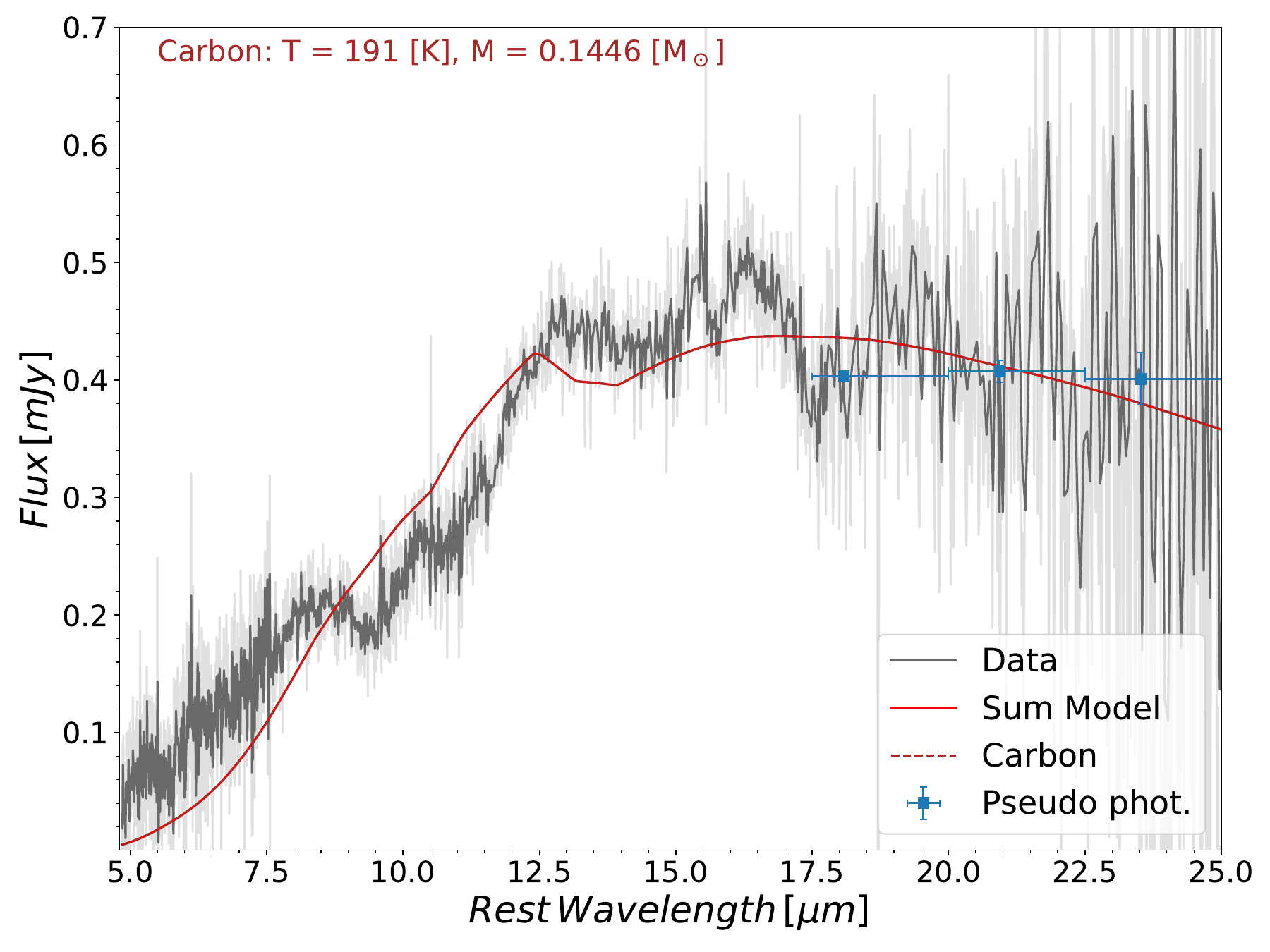}
    \includegraphics[width=0.45\textwidth]{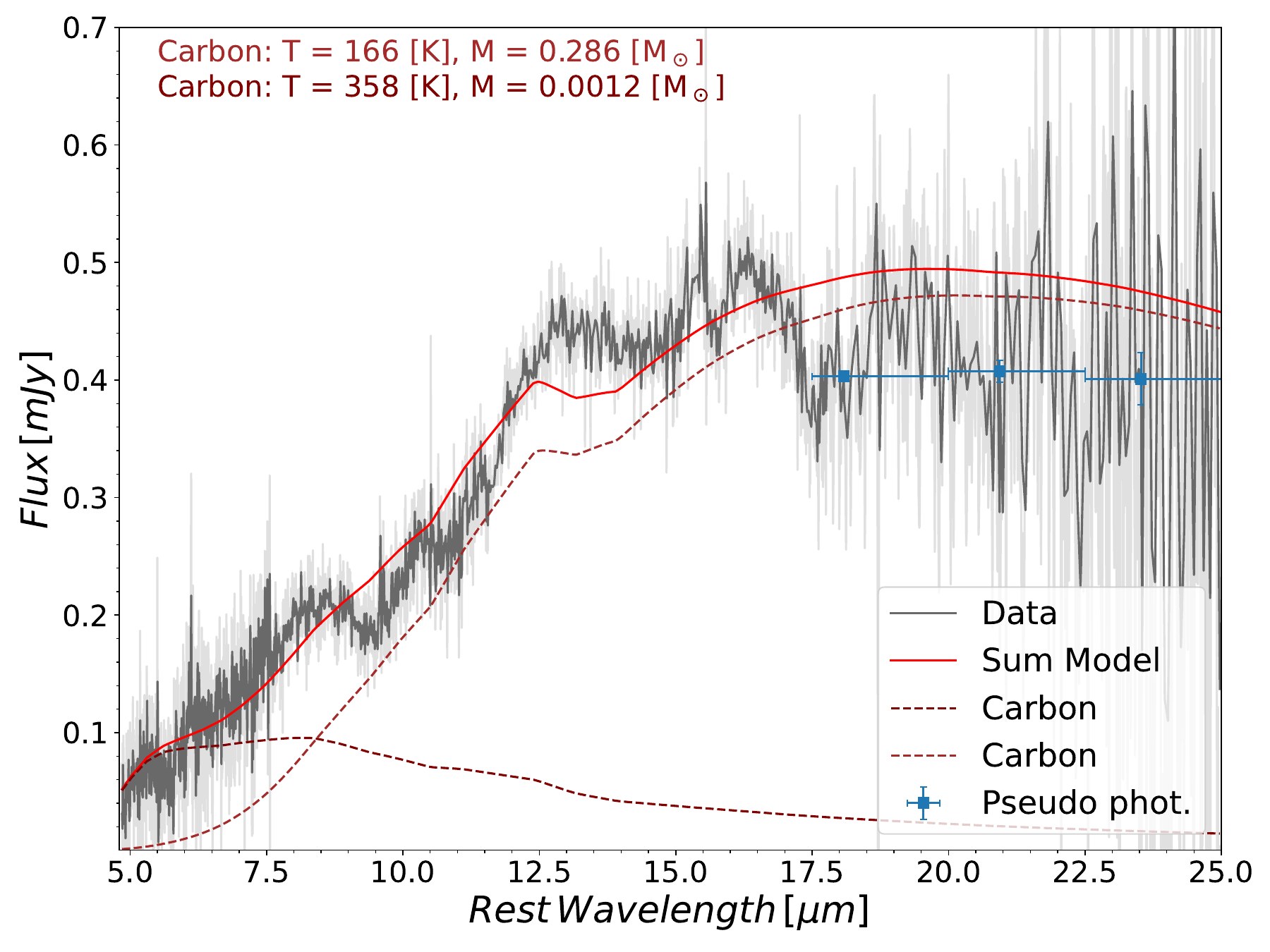}\\
        \includegraphics[width=0.45\textwidth]{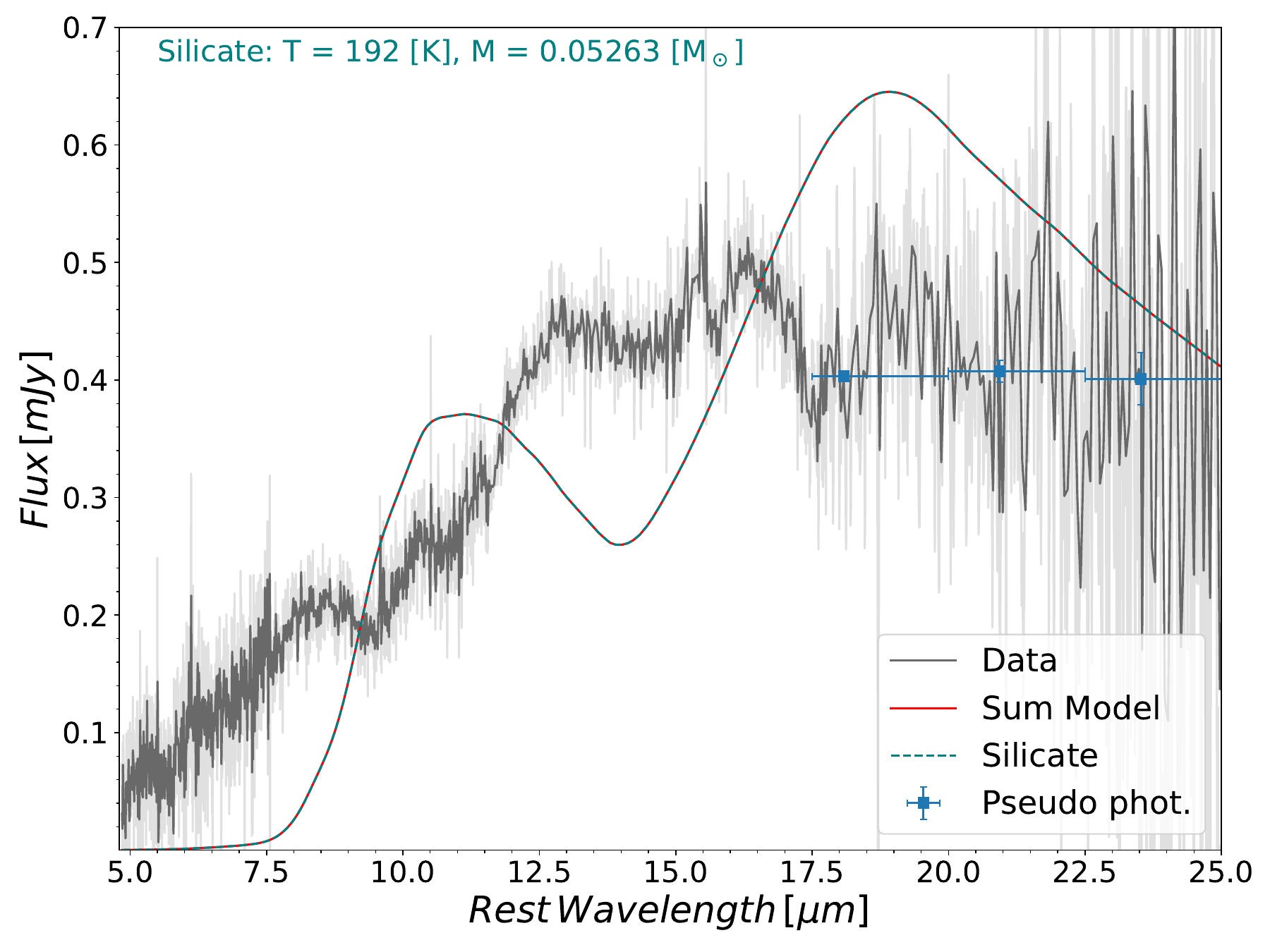}
        \includegraphics[width=0.45\textwidth]{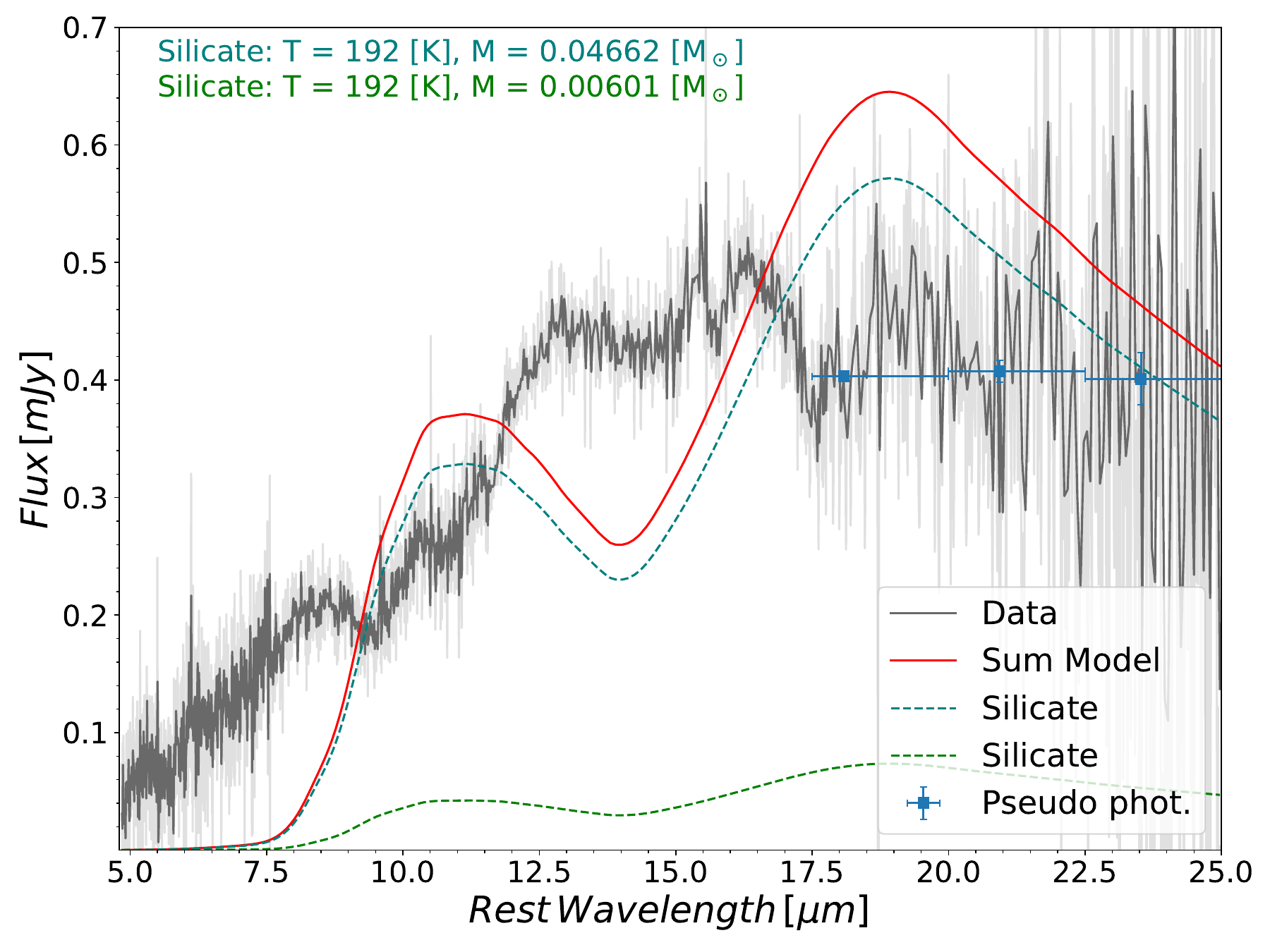}\\
           \includegraphics[width=0.45\textwidth]{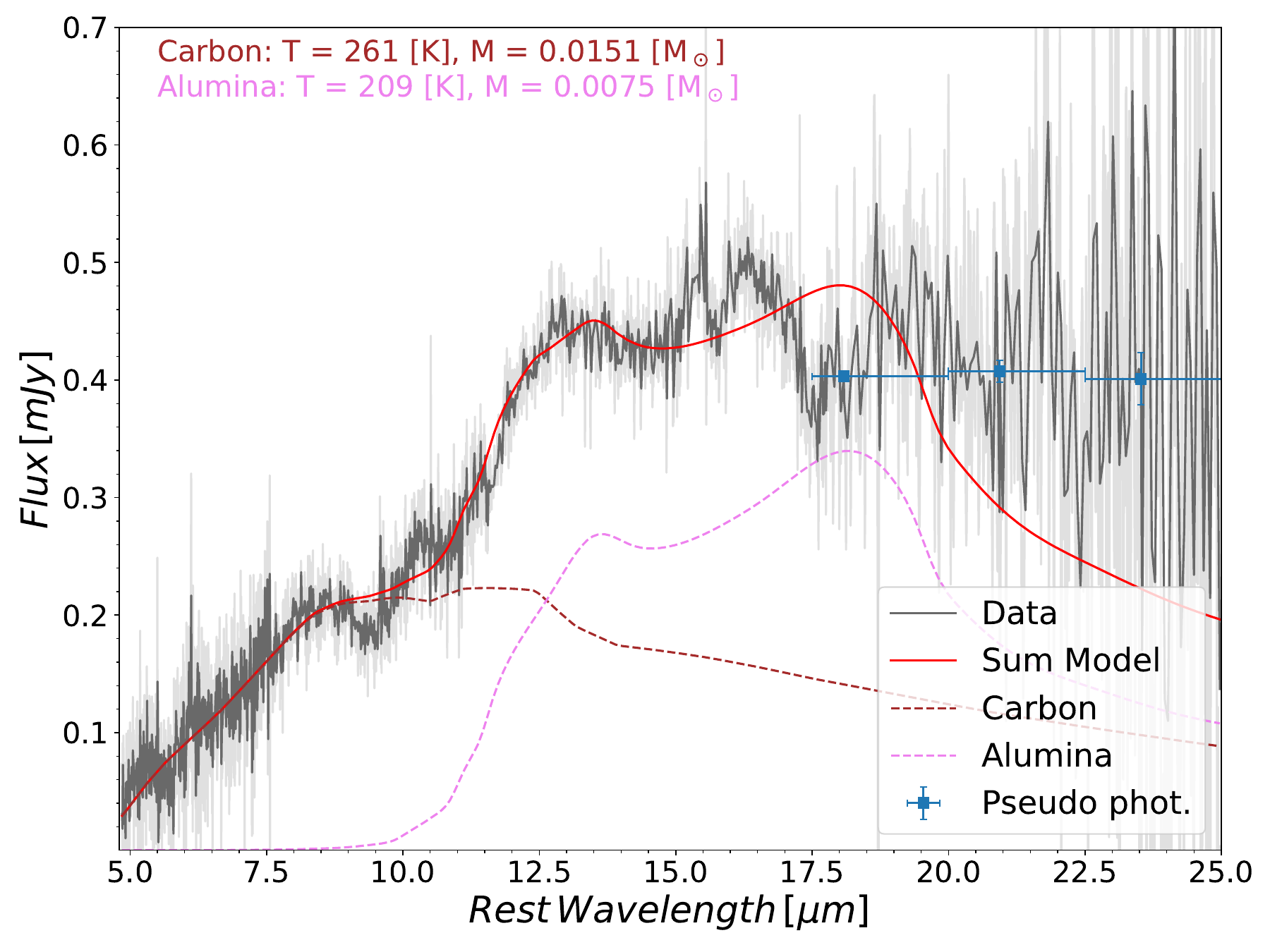}
            \includegraphics[width=0.45\textwidth]{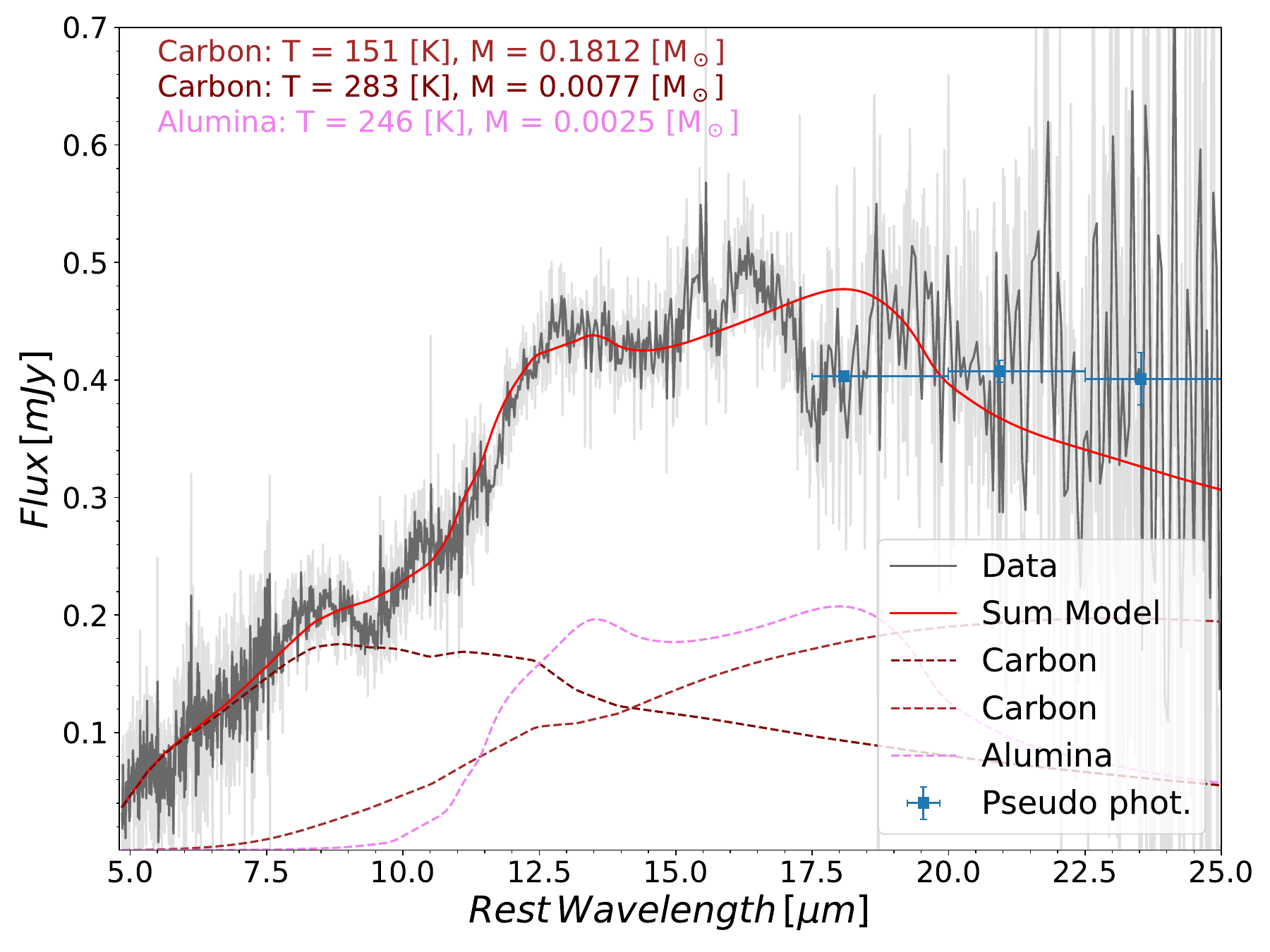}\\
    \caption{The MIR spectrum of SN 2010jl obtained with {\it JWST}/MIRI (light-gray) is the same as the black spectrum in Fig.~\ref{fig:miri}, while the darker gray is binned by 3 pixels) fitted with various optically thin emission dust models with several different models/assumptions as noted in each panel.  The blue points in each panel are pseudo-photometry made from large bins of the MIRI spectrum at $>$17 $\mu$m, with horizontal error bars corresponding to the bin widths.  For each panel, model SEDs of individual grain components are shown in dashed curves with colors noted in each legend, and the solid red curve shows the sum of the model SEDs (same as the individual component when there is only one). }
    \label{fig:all_models}
\end{figure*}

\subsection{Features in the Mid-IR Spectrum}

The mid-IR spectrum of SN~20210jl rises steadily from 5 to 17 $\mu$m, and then levels-off to longer wavelengths.  In the 5-17 $\mu$m portion, the steady rise is punctuated by a number of moderately broad features that may be a combination of emission and absorption features (see below).  At wavelengths beyond 17~$\mu$m, the noise level is quite high and it is unclear if any emission or absorption features are present. 

Broad features at 10-17 may be due to dust emission or absorption features, or a combination of both.  As described below, a likely explanation is silicate absorption features from a cool shell superposed on a blackbody continuum from warm optically thick dust.  A few narrow atomic emission lines are present in the spectrum, including [S~{\sc iv}] 10.4 $\mu$m and [Ne~{\sc iii}] 15.6~$\mu$m.  It is unclear if these are CSM intrinsic to the SN, or associated with unresolved H~{\sc ii} region emission.  However, [Ne~{\sc iii}] 15.6~$\mu$m does show a possible broad blueshifted component that is reminiscent of the H$\alpha$ profile discussed above, so a portion of this emission line's flux may arise from CSM interaction.
We do not detect strong polycyclic aromatic hydrocarbon (PAH) emission features in the {\it JWST} spectrum of SN~2010jl, although the usual PAH features at 7.7, 8.6, and 11.3 $\mu$m can be seen clearly in the extracted spectra of the two bright neighboring H~{\sc ii} regions A and B in Figure~\ref{fig:miri}.  PAH features were detected recently in similar {\it JWST} spectra of SN~2005ip \citep{shahbandeh24}.

\subsection{Modeling the MIR data} 
\label{sec:modeling}

Here we describe the modeling of the MIR spectra. Throughout this section, we assume that the MIR flux is dominated by thermal emission.   This thermal dust emission is either optically thin (Section 3.2.1), in which case the observed spectral energy distribution (SED) features are due to emission features from warm dust, or the warmest dust is optically thick (Section 3.2.2), in which case the SED features we observe are caused by a Planck function that gets partially absorbed by passing through cooler dust along the line of sight. We explore both options below.

\subsubsection{Optically thin dust emission}

To explore optically thin emission models, we fit the {\it JWST} spectrum using the equations and methods described by \citet{shahbandeh24}. 
Least-squares minimization was implemented using the Python \texttt{lmfit} package. 
We explored models with one, two, or three dust components varying in temperature, mass, optical depth, and composition, where the assumed composition determines the absorption coefficient, $\kappa$. Our models explore various dust species, including O-rich dust species like Mg-silicates, and C-rich dust species like amorphous carbon or graphite \citep{wesson21, ercolano07, sarangi13}. 
The absorption and emission characteristics of these grains are derived from \cite{draine07} for silicates and \cite{zubko04} for amorphous carbon; see  \citet{draine07} or \citet{sarangi22} for $\kappa$ values. 

The results from this optically thin modeling are shown in Figure~\ref{fig:all_models}, which includes examples of fits using different combinations of grain composition, including carbon (top row), silicates (middle row), and alumina + carbon (bottom row).  Of these three sets of models, optically thin silicate emission does the worst job of matching the spectrum, simply because the silicate emission peaks do not match local peaks in the observed spectrum.  Therefore, these models only illustrate that optically thin silicate emission cannot explain the observed spectrum, and the derived grain properties from these fits are not meaningful.  Based on data obtained about a decade earlier, \citet{wf15} also found the MIR spectrum to be inconsistent with showing a strong silicate emission feature.  The one-component and two-component carbon models  provide a better approximation of the spectrum than do silicates.  These capture the general flux level of the 5--15 $\mu$m SED well, but they do not adequately reproduce the detailed shape of features in the spectrum around 10, 12.5, and 16 $\mu$m.  The models with alumina + carbon are better still, but there is a peak at 17--18 $\mu$m in the model where a dip is seen in the observed spectrum, and the alumina models fall far short of the observed flux at long wavelengths (Fig.~\ref{fig:all_models}; bottom left) unless we add a larger mass of cool dust (Fig.~\ref{fig:all_models}; bottom right). 

Despite the generally poor fit to detailed features, this exercise is  useful for two reasons.  First, it indicates that optically thin emission with standard dust species is unlikely to account for SN~2010jl's MIR spectrum, so either higher optical depths (see below) or more complicated grain composition is needed.  Second, if the warm dust is in fact optically thick, then the dust masses derived from these optically thin models provide lower limits to the true dust mass.  For a model where a Planck function is absorbed by cooler dust (again, see below), there must be at least as much dust as in these optically thin models to produce the Planck function, but additional mass may of course be hidden inside the optically thick material, plus there would need to be additional dust at cooler temperatures to give rise to absorption features.  As such, lower limits to the dust mass are remarkable, since the values of 0.15--0.3 $M_{\odot}$ for the better models in Figure~\ref{fig:all_models} are already among the highest SN-produced dust mass estimated yet from MIR data.

\begin{figure}
    \centering
     \includegraphics[width=3.4in]{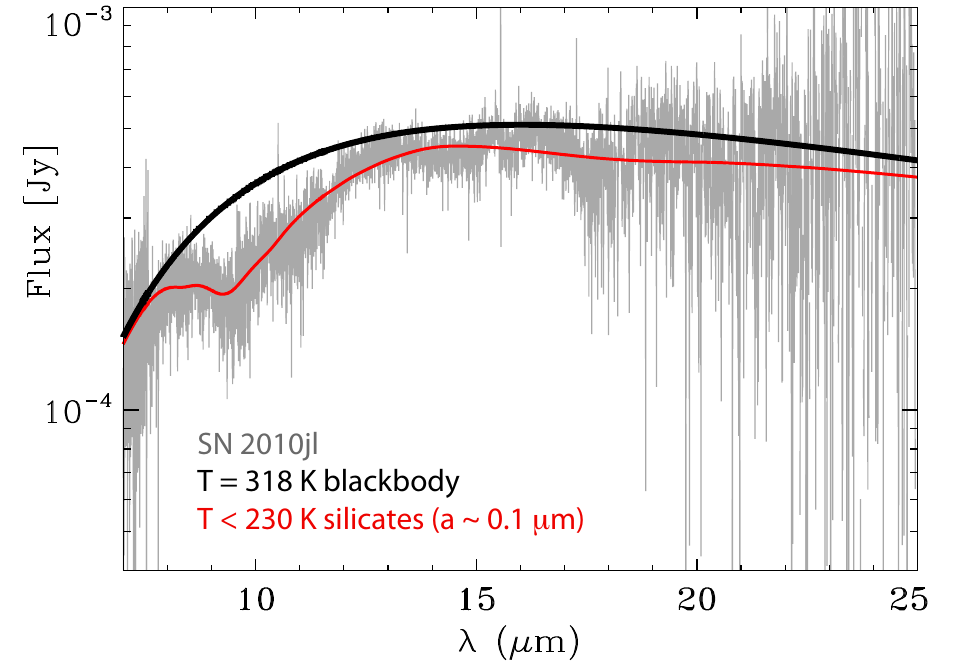}
    \caption{An alternative way to interpret the features in the observed {\it JWST} spectrum (gray), where a 318~K blackbody source (black) experiences absorption along the line of sight arising from cooler silicate dust (red), shown here with a maximum temperature of roughly 230~K and a grain size of 0.1 $\mu$m.}
    \label{fig:abs}
\end{figure}

\subsubsection{Dust emission and absorption}

In the previous section, the broad structures in the {\it JWST}/MIRI spectra were treated as optically thin dust emission features, with optically thin emission models yielding somewhat unsatisfying fits to the data.  Another possibility, however, is that the observed features in the MIRI spectrum arise from a combination of optically thick warm dust emission and optically thin absorption from significantly cooler dust located somewhere along the line of sight.  Figure~\ref{fig:abs} shows one example of such a model.  In this scenario, an optically thick background continuum source produces a warm $\sim$318~K blackbody spectrum (black curve), and then the cooler ($\lesssim$230~K in this example) foreground silicate dust absorbs in broad features at roughly 9.7 $\mu$m and 18--25 $\mu$m. In a separate paper, Dwek et al. (in prep.) explore this possibility in more detail, investigating fits to the shape of the observed {\it JWST} spectrum with various assumptions about the optical depth, temperature, and covering factor of the foreground silicate dust.  Here we only briefly consider the main implications.

In this scenario, the minimum (spherical) emitting radius for the 318~K blackbody source is about $R_{\rm BB} = 7 \times 10^{16}$ cm.  In order to emit a smooth blackbody, this warm dust source must be optically thick at all mid-IR wavelengths.  With $\tau \ge 1$, the minimum required dust mass is given by $M_d \ge 4 \pi R_{\rm BB}^2 / \kappa$.  Taking $\kappa$ at 7--8 $\mu$m (outside the silicate feature) to be about 300 cm$^2$ g$^{-1}$ for silicates \citep{sarangi22}, we have a minimum dust mass of $M_d > 0.1~ M_{\odot}$ for the warm blackbody-emitting source.  The true mass may, of course, be a great deal larger than this because more dust may be hidden inside the optically thick emitting geometry.  

Similarly, the mass of the cooler absorbing silicate dust depends on the radius where the dust is found.  The absorption near the peak of the 9.7 $\mu$m feature, where about 1/3 of the flux is absorbed, gives $\tau \approx 0.4$.  For a spherical shell, the mass of absorbing silicate dust is $M_d \approx 4 \pi R_{\rm abs}^2 \tau / \kappa$, where $\kappa$ at 9.7 $\mu$m is $\sim 3000$ cm$^2$ g$^{-1}$. Again, the radius $R_{\rm abs}$ of the absorbing dust is not known, but expressed in terms of the radius of the inner blackbody source, we have $M_d \ge 0.004 (R_{\rm abs}/R_{\rm BB})^2~M_{\odot}$.  $R_{\rm abs}$ must be significantly larger than $R_{\rm BB}$, because the dust must be cooler in order to be seen in absorption.  For a temperature of 230 K or less that is heated by the same source as the inner blackbody, we infer $(R_{\rm abs}/R_{\rm BB}) > 1.9$ (the absorbing dust may be at any larger radius, but it must be at least this far away to be cool enough).  This gives a total absorbing mass of $M_d > 0.014~ M_{\odot}$.  Together, the emitting and absorbing dust requires a minimum total mass of at least 0.11 $M_{\odot}$, but we reiterate that this is likely to be a significant underestimate.  It is unlikely that we have caught the warm emitting source at the exact moment just before it transitions from optically thick to optically thin; therefore, it is likely that $\tau$ is significantly larger than 1 for the hot emitting component, raising the dust mass of the warm component.  Additionally, nonspherical geometry or covering factors (clumpiness) for the absorbing shell will push the required absorbing mass up (see below).  We therefore consider it likely that a conservative estimate of the total dust mass in this scenario is around 0.2 $M_{\odot}$ or more, in agreement with the emission-only models above.  More precise values depend on various assumptions in the model fitting.  We consider an exploration of these to be beyond the scope of this work, and as noted above, we will investigate this scenario and the constraints on the total dust mass and detailed composition in more detail in a forthcoming paper (Dwek et al., in prep.).

The main difficulty with this scenario is coming up with a geometry that makes sense, since a spherical shell of cool absorbing silicates outside a central hot optically thick source cannot be easily reconciled with a spherical CSM interaction scenario for SN~2010jl.  The geometry must break spherical symmetry for this to work, and an interpretation of the geometry in this scenario of a warm emitting blackbody and cooler silicate absorption will be discussed below in Section 4.2.

\subsection{Total Dust Mass}

Regardless of the specific interpretation or the chemical composition of the dust, as discussed in the preceding subsections, it appears that a total dust mass of roughly 0.2 $M_{\odot}$ or more is required to account for the overall mid-IR emission detected from SN~2010jl at 13 yr post-explosion.  At a very basic level, this dust mass is physically plausible since the dust, if it is in the CDS, forms from shocked CSM at roughly solar composition.  The CSM, which arises from outer layers of the progenitor star that still contained significant H mass, is not heavily influenced by the inner O-rich SN ejecta; these layers are perhaps even CO-poor and N-rich.  Most evolved massive stars that still retain their H envelopes show significant N enrichment due to CNO burning products being mixed into the envelope.  Therefore, one might expect a standard gas:dust mass ratio of approximately 100:1.  In that case, the dust mass that we derive of order 0.2 $M_{\odot}$ is about 1\% of the total CSM mass of 10--20 $M_{\odot}$ needed to have been swept up in order to power the main light-curve peak \citep{ofek14,fransson14}.  In other words, SN~2010jl may have formed all the dust that it can from available refractory elements in the CDS.  If the dust mass continues to grow significantly, it may indicate that there is also dust formation in the inner SN ejecta, or that the CSM is more massive than earlier estimates.

\section{DISCUSSION}\label{sec:discussion}

\subsection{Diagnosing the Location of Emitting Dust}

There are three observational clues that are traditionally interpreted as new dust
formation\footnote{Here ``dust formation'' may also mean pre-existing
  grains in the CSM that were incompletely destroyed by the forward
  shock and then regrow.}  in SNe, as opposed to IR radiation arising from pre-existing dust in the CSM that is still exterior to the forward shock.  These are (1) a growing IR excess
consistent with dust emission, (2) an increased rate of fading in the
optical continuum  attributed to increased extinction from
newly formed dust, and (3) a progressively increasing blueshift in
emission-line profiles caused by dust within the line-forming region that preferentially obscures
redshifted portions of the explosion.  When all three occur simultaneously, as in the case of SN~1987A
\citep{danziger89,lucy89,gn87,gn89,wooden93,colgan94}, this gives a strong
indication that new dust has formed.

SN~2010jl showed all three of these signposts \citep{kc17}.  (1) It has strong mid-IR excess at many epochs \citep{andrews11,gall14,wf15,kc17,sarangi18}.  (2) Its visual-wavelength light curve has an inflection $\sim 300$ days post-explosion, where the decline steepened \citep{fransson14,jencson16,kc17}.  This rapid drop in visual light was accompanied by an increase in the thermal-IR flux, suggesting that the bolometric luminosity maintained a continual slow decline as more of the flux shifted to the IR \citep{sarangi18}.  (3) Beginning shortly after peak brightness, IW emission lines have shown a deficit of flux on their red wings, starting as a subtle asymmetry and then growing to a pronounced blueshifted peak \citep{smith12,gall14}, and eventually removing all of the redshifted emission (this work).

Unraveling the details of new dust formation in SNe can be a tricky business, however. Alone, each of these three clues can be somewhat ambiguous. Clue \#1 is ambiguous in the sense that an IR excess could also arise from an IR echo from pre-existing dust, as noted above, but the time evolution of the SED can help.  Clue \#2 only works if we know the intrinsic rate of fading before extinction alters the light curve (or if we have the full optical-to-IR SED; see below). While this is plausibly known for the radioactive-decay tail of normal SNe, the rate of fading in interacting SNe can follow almost any slope depending on the density profile of the CSM, and it may change suddenly as the shock encounters local CSM density fluctuations.  This leaves Clue \#3 as arguably the least ambiguous signpost of new dust formation, since dust growth must occur within the expanding SN itself in order to influence line profiles.  Even this is not without caveats, since some other effects can also cause blueshifted profiles (such as occultation of the back side of the SN by the SN photosphere, which, however, only works at early times; \citealt{smith12}).
Helping to counteract ambiguity is that Clues \#2 and \#3 are also wavelength dependent, since they arise from extinction caused by the newly formed dust; dust typically causes more extinction at shorter wavelengths, as long as the grain size is not much larger than the observed wavelengths.  This wavelength dependence has been observed in SNe~IIn, although it usually suggests large grains \citep{gall14,smith12,smith20,smith23da}, at least compared to the very small grains typically found in the ISM.  Overall, when asymmetric blueshifted emission-line profiles persist to late times, when this line-profile effect is wavelength dependent, and when this is accompanied by an IR excess that can be explained by dust emission, this is usually taken as a strong indication of new dust formation.  

Since extragalactic SNe (with the exception of SN~1987A) are spatially unresolved at visual and IR wavelengths, the precise location of the dust is often a stubborn mystery to solve.  This is especially true in strongly interacting SNe, because these have an additional possibility that normal SNe do not have.  An important feature of strongly interacting SNe is the formation of a CDS between the forward and reverse shocks
\citep{chugai01,chugai04,smith08tf}.  Efficient radiative cooling
causes the dense post-shock gas to collapse into a thin, dense, clumpy,
and probably well-mixed layer \citep{vanmarle10}.  This rapid cooling
that forms a dense shell is a unique feature of interacting SNe,
causing their defining narrow-line spectra and their high
luminosity.  

This rapid cooling and high density in the CDS may also
trigger early dust formation.  Thus, for interacting SNe, we have three possibilities for the location of dust that may give rise to IR excess emission:  (1) pre-existing CSM dust (observed as an IR echo; \citealt{andrews11,gerardy02}), (2) newly formed dust in the freely expanding SN ejecta (best traced by mid/far-IR excess emission), and (3)  newly formed dust in the compressed and radiatively cooled post-shock regions (i.e., the CDS), which is unique to interacting SNe.  These options for the
location of the dust are not mutually exclusive, and in fact, they are correlated.  In order to make a SN~IIn, very dense CSM is needed to cause strong post-shock compression, and this same dense CSM is likely to be dusty and thus give rise to an IR echo.  Some massive-star progenitors of CCSNe can have strong mass loss before the SN, and known types of progenitor stars that might provide this CSM are either extreme dusty red supergiants (RSGs) like VY~CMa \citep{smith01,smith09} or luminous blue variables (LBVs; \citealt{smith26}).   Truly dust enshrouded RSGs are, however, extremely rare \citep{bs22}, compared to normal-winded RSGs \citep{beasor20}.  LBVs tend to be surrounded by dusty winds or shells \citep{so06,smith26}, but these stars are also extremely rare.  In both cases, the dust-enshrouded phase is very short-lived, requiring an extremely rare type of star to have its strong mass loss synchronized with explosion.  This is likely why SLSNe IIn are so rare. 

The physical cause of this pre-SN mass loss remains an enduring puzzle.  With such massive and dusty CSM, thermal-IR radiation from the IR echo may persist at some level, even as the forward shock sweeps through the CSM and simultaneously forms new dust in the post-shock CDS.   Moreover, new dust formation in the post-shock gas does not preclude dust formation in the inner SN ejecta, so any interacting SN may have dust in all three zones.  Each may contribute some of the IR emission at a particular wavelength, and these relative contributions may evolve over time. For SNe IIn and especially superluminous SNe IIn, one might naturally expect a sequence where the IR excess at early times is dominated by an IR echo, and then the MIR contribution from the CSM echo diminishes as the forward shock sweeps trough the CSM and destroys CSM dust, while the MIR contribution from post-shock and ejecta dust grows steadily with time.  Essentially, the CSM dust moves from the unshocked CSM to the post-shock CDS, and some unknown fraction of this dust may be destroyed and then regrows in the process.  Regrowth of this incompletely destroyed dust, as opposed to nucleating grains from the gas phase, may help explain why strongly interacting SNe seem to form post-shock dust so rapidly \citep{smith08jc}.  This post-shock dust is also warmer (as compared to SN ejecta dust at a comparable late phase) and therefore easier to detect in the mid-IR.

\begin{figure*}
    \centering
    \includegraphics[width=5.0in]{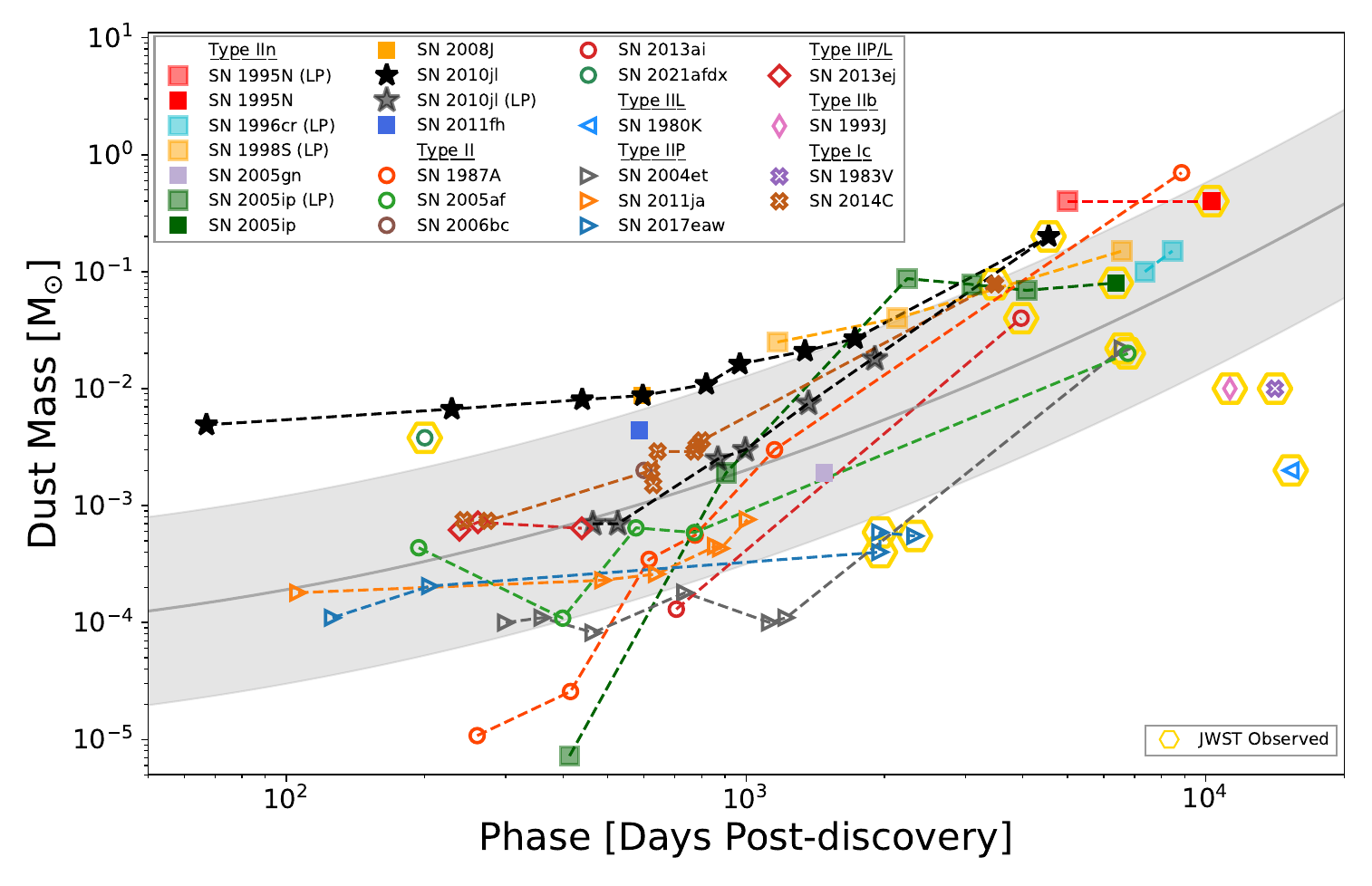}
    \caption{The dust mass in SN~2010jl as a function of the epoch of MIR observations compared to other SNe with estimated dust masses in the literature. Most of these come from MIR observations with {\it Spitzer} and {\it WISE}, or with {\it JWST} (outlined by gold hexagonal segments). A few were estimated from blueshifted line profiles (LP), or in the special case of SN~1987A, from a variety of observations including far-IR/submm data.  Overall, the inferred dust mass in SN~2010jl is one of the highest at all epochs.  For SN~2010jl, we show both near/mid-IR mass estimates from  \citep{szalai19} and this work in black stars, while dust masses estimated from blueshifted emission-line profiles (LP) are shown with dark gray stars \citep{gall14,bevan20,nicu22}.  LP estimates only trace new dust, whereas mid-IR emission includes both new dust and pre-existing CSM dust. As discussed in the text, SN~2010jl's large dust mass at early times may be mostly CSM dust, while at later times the dust mass is dominated by newly formed dust in the post-shock CDS and ejecta.  LP estimates are also shown for SN~2005ip and SN~1995N.  Most objects are MIR estimates from the compilation by \citep{szalai19}, except SNe~2005af, 2017eaw, and 1980K \citep{zsiros24,pearson25,sarangi25}, and the large dust mass in SN~1995N reported recently by \citep{clayton25}.}
    \label{fig:masstime}
\end{figure*}

Examining Figure~\ref{fig:masstime}, we see that the MIR-derived dust mass of SN~2010jl is on the high end at all epochs, compared to MIR-derived dust-mass estimates for many SNe published in the literature.  An unusually high mass of CSM is required to power the main light curve of SN~2010jl (even compared to most SNe~IIn), and the progenitor star is thought to have been heavily dust enshrouded \citep{dwek21}.  Therefore, we suggest that a scenario like the one described above likely explains the evolution of the inferred dust mass in SN~2010jl.  Namely, for the first few hundred days, the MIR-derived dust mass is probably dominated by an echo from CSM dust, even as post-shock dust formation begins.  But as time passes, an increasing fraction of the CSM dust is swept up and destroyed (perhaps incompletely) by the forward shock, and simultaneously, new dust forms (or the grains rapidly regrow) in the post-shock zone, gradually taking over the MIR emission by day $\sim$1000.  The inner SN ejecta dust may also form new dust as time proceeds and may also produce a large mass of dust, but the bulk of the MIR emission we detect with {\it JWST} may still be dominated by the warmer post-shock dust at late times, since it is continually heated by lingering shock interaction.

In the case of SN~2010jl, visual-wavelength spectra obtained at very late times, 7--11 yr post-explosion (see Fig.~\ref{fig:vel}), now confirm earlier claims \citep{smith12,gall14,sarangi18} that a significant mass of dust has formed in the post-shock CDS region of SN~2010jl.  The fact that the blueshift has grown with time and that it has persisted for a decade after explosion, despite the fading of the visual-wavelength continuum luminosity, rules out any non-dust explanation for the origin of the blueshifted asymmetric IW line profiles.  Other potential mechanisms such as occultation by the SN photosphere \citep{smith12,dessart15} and electron-scattering effects \citep{fransson14} rely on high continuum optical depths in ionized gas in order for electron scattering to work, and this is impossible at such late phases when the SN has cooled and faded.  Indeed, the IW emission lines in the spectrum have lost their broad electron-scattering wings that characterized the spectrum at early times.  Once dust forms, on the other hand, the extinction remains even as the material cools.  In fact, additional grain growth is likely to occur in the CDS as material cools further, consistent with the observed trend that the blueshift becomes even more pronounced with time.  Even if the formation of new dust grains stops at some late epoch, it is possible that the blueshifted line profiles can persist, unless grains get destroyed in the CDS.

Of course, from available information in this unresolved extragalactic SN, we cannot rule out the possibilities that there is also some dust in the pre-shock CSM or some newly formed dust in the inner SN ejecta.  But the IW line-profile evolution in spectra demand that there is some new dust in the post-shock CDS layer of SN~2010jl.  In addition, it is worth noting that from data presented here and in other papers, we cannot confidently distinguish between the two alternatives of (1) new dust grains that have condensed from the gas phase in the post-shock CDS, as opposed to (2) dust grains that were pre-existing in the CSM, but were incompletely destroyed by the forward shock, only to rapidly regrow in the dense post-shock layer.  Either scenario is likely to cause an apparent increase in the amount of dust in the post-shock CDS, satisfying observational constraints of an increased IR excess, strengthening blueshift in IW line profiles, and increased rate of fading in the light curve. 

It is worth noting, however, that a significant contribution from very hot dust at 1000--2000 K, which would correspond to CSM dust that is heated by the shock to the point of vaporization, is not clearly seen in the late-time data.  As noted above, there is a possible hot 2800 K continuum source seen in the visual-wavelength spectrum, but this may be largely due to contamination from nearby RSGs.  In any case, the total luminosity of any such hot dust is less than about 1\% of the 200--300~K dust component we detect in the MIR with {\it JWST}. This seems to suggest that at the current epoch, there is not a great deal of CSM dust being overtaken by the forward shock; it may have been important at earlier epochs, as noted above.  This also implies that the cooler absorbing dust required for our emission + absorption model for the SED is likely to be at a location that is detached from the CDS in order to prevent the emission from dust at all temperatures between 300 K and 200 K.

\begin{figure}
    \centering
    \includegraphics[width=0.42\textwidth]{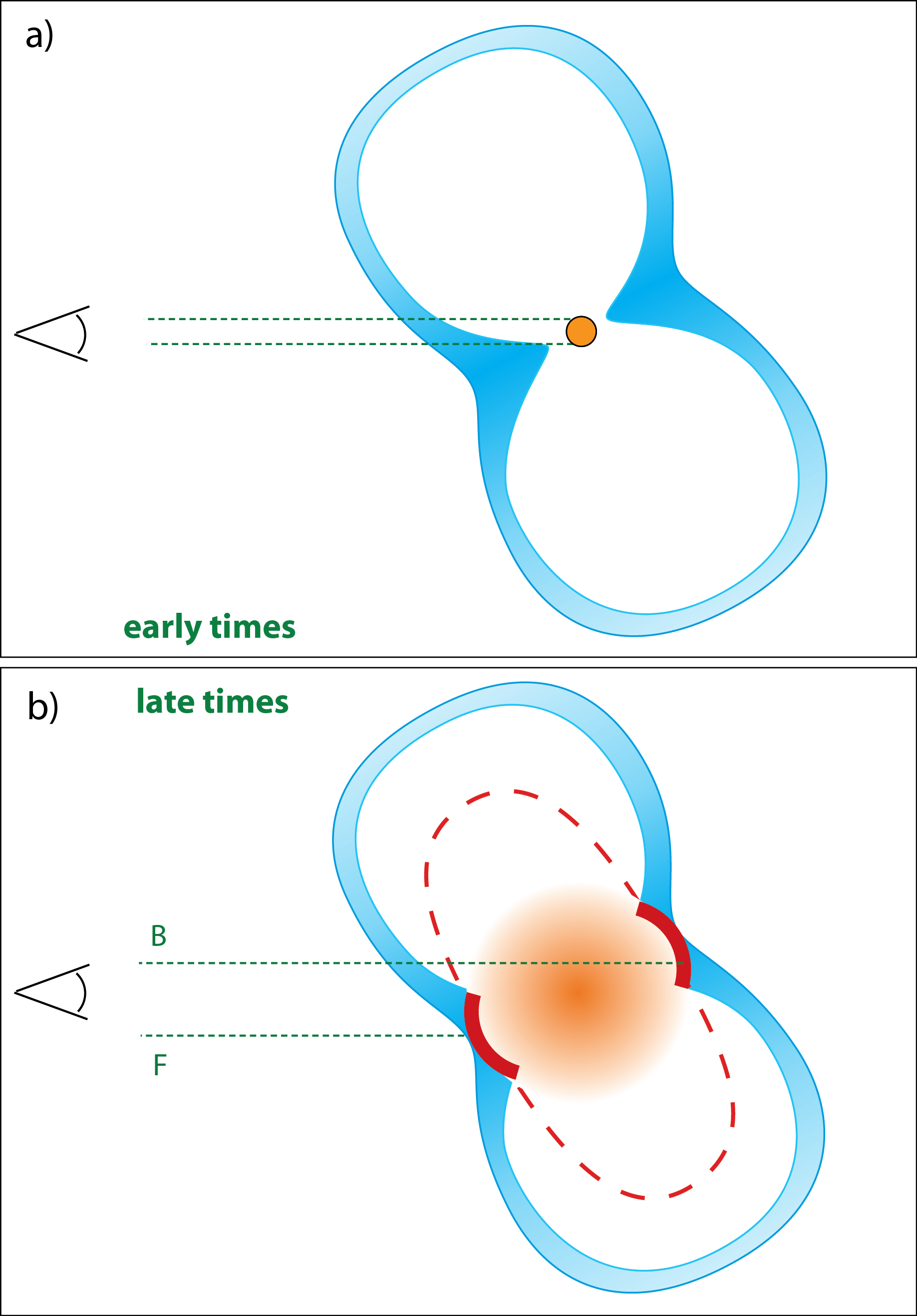}
    \caption{A sketch showing a side view of a possible CSM interaction geometry at early times during the main light-curve peak (top panel $a$), and at a late epoch appropriate to the {\it JWST} observations reported here (bottom panel $b$).  The blue feature is a pre-exiting bipolar CSM shell, the thick red arcs are the main CSM interaction shock in the equatorial region, and the red long-dash curves represent the forward shock in the polar directions, which has reached a larger radius than in the equatorial region because of the lower CSM density there.  The orange inner region is the SN photosphere (top panel) or inner SN ejecta (bottom panel).  In the lower panel, this orange region is approximately the size of the dust-formation region assuming that ejecta dust forms out to velocities of $\sim 2650$ km s$^{-1}$ \citep{sarangi22}. The dashed green horizontal lines represent lines of sight through the CSM to the SN ejecta (top panel) or to the front (F) or back (B) sides of the CSM interaction region (bottom).}
    \label{fig:geom}
\end{figure}

\subsection{Diagnosing the Location of Possible Absorbing Dust}

As discussed above in Section 3.2.2, a possible interpretation of the {\it JWST}/MIRI spectrum is that the features we observe arise from a combination of a warm 318~K blackbody and cooler silicates that cause absorption (see Figure~\ref{fig:abs}).
The main challenge for this model is understanding the implied geometry, because a straightforward spherical geometry poses problems.

At such late phases around 13 yr after explosion, any heating of dust by radioactivity in the inner ejecta is minimal.  Dust located in the inner SN ejecta would be cold and its emission would be observable only at far-IR or submm wavelengths, as in the case of SN~1987A \citep{matsuura15}.  The warmest dust we observe (the 318~K component in this scenario) must be located closest to the primary source of luminosity, which at these late phases is the CSM interaction shock (i.e., the main forward shock).  It is therefore very likely that the 318~K component would be found in the CDS itself, heated by the shock, since we know that there is a significant amount of dust located there based on the asymmetric line profiles seen at visual wavelengths, as discussed above.  The radius of the CDS at such a late phase must be at least $R_{\rm CDS} = v \times t$.  For $v_{\rm shock} \approx 2500$ km s$^{-1}$ (from Figure~\ref{fig:vel}) and $t=13$ yr, we have a radius of $\sim 9.7 \times 10^{16}$ cm.  This is certainly a lower limit, since $v_{\rm shock}$ was faster at earlier epochs.  A radius of 10$^{17}$ cm is marginally larger than the value of $R_{\rm BB}$ noted above for the 318~K component, which makes sense if the geometry is not spherical (see below).  

A difficulty arises, then, in deciding where to place the cooler absorbing component.  In a spherical model, the cooler absorbing dust cannot be from SN ejecta, because it would be inside the optically thick region.  If it were CSM dust, it would need to be at a significantly larger radius in a detached shell.  If the dust were found in a continuous wind ahead of the shock, then there would be a continuous range of dust temperatures in the pre-shock zone from 318~K and down, and this emission would fill in the absorption or even give rise to emission features.  Therefore, in a model where pre-shock CSM gives rise to these absorption features, there would need to be a significant gap between the warmest dust in the post-shock zone and the cooler dust at larger radii.  This is implausible as well, however.  Any dust that has $\tau=0.4$ at 10 $\mu$m would have a very large visual-wavelength optical depth along our line of sight to the SN.  If this is pre-existing dust in the CSM that was evidently not destoyed, it should have been present at early epochs and would have caused significant extinction and reddening of the light of the SN itself.  After correcting for a small amount of foreground extinction from the Milky Way and host galaxy of $E(B-V)\la0.05$ mag, UV/optical photometry during the main light-curve peak indicated continuum temperatures around 6000--7000 K \citep{fransson14}, which is typical of SN~IIn photospheres and rules out a large amount of circumstellar extinction at early times. 

These complications might be mitigated if we allow the CSM geometry to be highly nonspherical.  Consider the bipolar geometry shown in Figure~\ref{fig:geom}, which depicts the CSM interaction geometry at early times (panel $a$) during the main peak of the SN light curve, and at the current late epoch (panel $b$).  A similar bipolar geometry viewed from a low- or mid-latitude vantage point has been suggested previously based on the different velocities of narrow H$\alpha$ absorption and emission in early-time spectra \citep{smith11}, polarization of SN~2010jl \citep{patat11}, and also polarization of a sample of SNe IIn in general \citep{bilinski23}.  In this picture, the warm 318~K dust that produces the bright mid-IR blackbody emission is located predominantly in the shock that runs into dense CSM at the pinched waist of the bipolar shell, near the equatorial regions (colored red in Figure~\ref{fig:geom}b).  

With this geometry, there are two potential locations for the cooler absorbing dust: (1) in the partly detached walls of the bipolar lobes in the CSM (blue in Fig.~\ref{fig:geom}), or (2) in the inner SN ejecta (orange in Fig.~\ref{fig:geom}).
Emission from the far side of the shocked torus would pass through the near wall of the bipolar shell (blue) on its way to the observer, and the cooler dust in the walls of this bipolar shell (separated from the warm dust emission source and at a larger radius) might give rise to the silicate absorption features we see in the MIRI spectrum (this is along line of sight ``B" tracing emission from the back of the shocked torus in Fig.~\ref{fig:geom}).  Emission from the front of the shocked torus (line of sight labeled ``F" in Fig.~\ref{fig:geom}) may pass through a smaller column of warmer dust in the unshocked CSM, giving rise to less silicate absorption and potentially avoiding a large amount of emission from hot dust immediately ahead of the shock.  This scenario has a remaining difficulty, though: if the CSM dust has a sufficiently high column density to cause strong MIR silicate absorption at this late epoch, it is difficult to avoid having this same dust absorb much of the UV and visual-wavelength light during the main peak at early times (i.e., if the CSM dust is still present now, then it was not destroyed by the SN radiation and should also be along the line of sight at early times).  As noted earlier, however, the line-of-sight reddening to the SN was low.  

Examining Figure~\ref{fig:geom}, an intriguing alternate possibility is that the silicate absorption features arise in the cooler central SN ejecta (the inner orange gradient region in Fig.~\ref{fig:geom}), since the ``B" line of sight may also pass through the inner freely expanding SN ejecta on its way to the observer, provided that the SN is observed at a moderately high inclination angle.  As noted earlier, the expansion speed of the main forward shock in the equatorial regions is $\sim 2500$ km s$^{-1}$.  Coincidentally, this is similar to the velocity of 2650 km s$^{-1}$,  inside of which dust is most likely to form\footnote{This represents the velocity coordinate for the outer edge of the He core in SN~1987A, marking its main dust-formation zone \citep{sarangi22}.  Obviously this value might be somewhat different in SN~2010jl, but this is a useful reference.} in SN ejecta \citep{sarangi22}. Thus, SN ejecta dust may fill most of the volume interior to the main shock front in the equatorial region (i.e., most of the orange region in Figure~\ref{fig:geom}b).  This avoids the problem of the previous case where the MIR absorbing component should also have absorbed the SN light at early times, because in this alternate case, the SN ejecta dust had not yet formed at early times and the ejecta are inside the SN photosphere anyway.    Obviously, silicate absorption might occur in both the central SN ejecta and the near side of the bipolar CSM shell, but this second scenario allows one to have a strong 10 $\mu$m silicate absorption feature from cold dust present at late times that was not there at early epochs.  For this reason, we prefer the option where a majority of the silicate absorption occurs in the cool inner SN ejecta, whereas the MIR emission from warm dust arises from dust in the CDS heated by CSM interaction.   In either case, the silicate absorption feature along the line of sight to the far side of the torus would be deeper than only 1/3 of the flux if it could be spatially resolved, but its strength is diluted by the contribution from the near side of the torus when both are included in the same unresolved source.  When modeling this absorption, this dilution by the unabsorbed near-side emission might be accounted for by using a clumping factor or by adopting a temperature for the absorbing dust that is higher than the actual temperature.

In this scenario, the ability to detect the SN ejecta in absorption depends on viewing the event from a relatively high inclination angle (roughly $i \ga 50^{\circ}$) so that the light from the far side of the equatorial CSM interaction region passes through enough of the interior ejecta.  Other SNe~IIn may not show this type of ejecta absorption if they are viewed at moderate inclinations or nearly pole-on.  The SN must also be observed at a very late epoch, so that the dusty portions of the SN ejecta have enough time to expand to a large radius and become optically thin.  This may explain why this sort of silicate absorption feature is rarely seen in SNe~IIn.  Note that in this scenario, the high optical depth of the warm emitting component (essentially a Planck function) means that the silicate emission feature is suppressed.  Of course, it would be a mistake to interpret this lack of silicate emission as indicating carbon-rich grains in the CSM interaction region or CSM. 

\subsection{Total Dust Mass Formed by SN~2010jl Over Time}

From the {\it JWST}/MIRI spectrum obtained on 2023 April 19 (4550 days $\approx$ 12.5 yr post explosion), we have estimated an emitting dust mass of roughly 0.2 $M_{\odot}$ in SN~2010jl (with a lower limit of 0.11 $M_{\odot}$).   This is among the largest dust masses measured for an extragalactic SN explosion, significantly exceeded only by the larger dust masses estimated for SN~1987A and Cas A.   Of course, these objects are older and their large dust masses derived from observations at longer wavelengths; moreover, the dust mass in SN~2010jl may continue to grow (although that additional dust may also end up being cooler and not easily detected at MIR wavelengths). Considering dust masses of extragalactic SNe estimated from MIR data, it is noteworthy that the three largest dust masses that have been measured have been for SNe~IIn, including almost 0.1 M$_{\odot}$ in SN~2005ip \citep{shahbandeh24}, at least 0.2 M$_{\odot}$ in SN~2010jl (this work), and $\sim 0.4$ $M_{\odot}$ in SN~1995N \citep{clayton25}.  It is likely that continued CSM interaction keeps the dust warmer in SNe~IIn as compared to normal SNe, facilitating its continued late-time detection at MIR wavelengths.

Our derived dust mass at almost 13 yr post-explosion is $\sim$80 times larger than the dust mass of $2.5 \times 10^{-3}$ $M_{\odot}$ estimated a decade earlier from shorter-wavelength data taken 868 days post-explosion \citep{gall14}, and about 10 times larger than the value estimated on day 1909 \citep{nicu22}.  This indicates the substantial growth of new dust over time, or a substantial drop in optical depth.  Most of this dust must be newly formed dust; if a significant fraction of the currently emitting dust were CSM dust, it should also have been there a decade ago, but observations in 2012 reveal a much smaller mass of emitting dust.  Interestingly, the growth rate in the dust mass is quite consistent with expectations from a comparison to other SNe, and in very good agreement with the expected growth of SN~2010jl's dust mass shown by \citet{gall14}.  Figure~\ref{fig:masstime} compares the recent dust mass measurement of SN~2010jl with several other objects having measured dust masses, including late-time observations of SN~1987A.  We see here that SN~2010jl matches the general trend of increasing dust mass with time (see Figure~\ref{fig:masstime}), and is comparable to teh trajectory of SN~1987A.  SN~2010jl has a consistently high dust mass compared to most other SNe at each epoch.

\subsection{Post-Shock Dust Formation in SNe IIn}

As noted in Section~1, dust seen in high-redshift galaxies requires an early source of dust-grain production in the Universe, before AGB stars and other sources from the evolution of low-mass stars become available.  CCSNe from massive stars are a viable candidate for the source of this dust, if they can supply 0.1--1 $M_{\odot}$ per SN event.  From recent {\it JWST} observations, we find that at almost 13 yr post-explosion, SN~2010jl has already produced a dust mass of at least 0.11~$M_{\odot}$ and probably more than 0.2 $M_{\odot}$, and from a comparison with SN~1987A and other SNe (see Fig.~\ref{fig:masstime}), SN~2010jl appears to be on track to form even more (perhaps $\sim 1$~$M_{\odot}$, as in SN~1987A).  The additional dust that may appear at later times, however, is likely to be SN ejecta dust, since the dust formed from shocked CSM is likely to be limited to about 1\% of the total CSM gas mass.  The direct implications for dust at high redshift are difficult to infer, however, since we do not know how representative SN~2010jl may be.  In the local Universe, SNe~IIn represent only 10\% or less of core-collapse SNe \citep{smith11frac}, and superluminous events are a small fraction of those.  On the other hand, SNe IIn and especially SLSNe IIn seem to prefer low-metallicity dwarf host galaxies \citep{moriya23,stoll11}, so they may be influential if they are more common in the early Universe.

Another important consideration is not only the initial {\it production} of new dust grains by a CCSN, but also the {\it survival} of that dust.  In a traditional CCSN without strong CSM interaction, the main site for dust formation is in the inner metal-rich SN ejecta.  This is the case for the large masses of dust detected in SN~1987A and Cas A.  But as noted earlier in the Introduction, an unknown fraction of this dust might never make it to the ISM because it may be destroyed as it passes through the reverse shock \citep{bs07,micelotta16,bc16}.  This is likely to temper the contribution by normal SNe to the ISM dust budget. 
Similarly, a significant fraction of the CSM dust that may have formed in the cool wind of the progenitor may also be destroyed by the fast forward shock in normal SNe. 
The fraction of SN dust that survives passage through the reverse shock evolves with time (i.e., the conditions of the reverse shock change with time as the remnant evolves), and it depends on the initial grain size, with larger grains being more resilient.  In a recent investigation of these conditions in the Cas A remnant (including estimates of the remnant's future evolution), \citet{k24} find that between 17\% and 28\% of the ejecta dust grains are ultimately able to survive passage through the reverse shock.  This suggests that a majority of the 0.5--1 $M_{\odot}$ of dust produced in the ejecta of SNe, like Cas A and SN~1987A, will be destroyed.

The situation is very different in SNe~IIn and SLSNe~IIn that show dust formation in the post-shock CDS, because in these cases, the new dust that is observed to form is already in the post-shock region.  This dust is therefore not in danger of destruction by either the forward shock or the reverse shock, and it is therefore likely that most of it will ultimately make it to the surrounding ISM in the host galaxy.  Another consideration for the dust survival is the grain size.  In superluminous events like SN~2010jl \citep{smith12,gall14} and SN~2017hcc \citep{smith20}, the wavelength dependence of the blueshifted line asymmetry has indicated $R_V$ values of 6--10, significantly larger than the average Milky Way ISM value of $R_V=3.1$.  This suggests that the new dust grains in the post-shock CDS region are larger than grains in the normal ISM.  Following the study of Cas A \citep{k24}, these SN IIn grains will be more resilient if they do encounter shocks or processing by UV radiation.  Lastly, SNe~IIn and SLSNe~IIn have heavy mass-loaded shocks that decelerate quickly, having given up most of their kinetic energy to power the bright SN light curve in the year after explosion \citep{sm07}.  These slower shocks may also be more conducive to continued dust formation and survival. 

In addition to the dust in the post-shock CDS, SNe IIn may also, of course, form dust in their inner ejecta, just like normal CCSNe. With a slower shock front, this ejecta dust might also be more likely to survive.  Our understanding of the dust contribution by SNe IIn and SLSNe IIn would benefit from a more detailed consideration of these points.  Nevertheless, there appear to be good reasons to think that SNe~IIn are not only efficient dust producers, but that the dust they make is more likely to survive than the ejecta dust in normal SNe.  These considerations might help make the dust contribution by SNe~IIn competitive with that of normal core-collapse events, despite SNe~IIn being a smaller fraction of all CCSNe. For example, even considering the observed SN subtype fractions \citep{smith11} where SNe~IIP are about half of all CCSNe and SNe~IIn are only about 10\%, their dust contribution to the ISM would be the same if only 20\% of the SN IIP dust survives and all of the SN~IIn dust survives.

\section{SUMMARY AND CONCLUSIONS}

We presented new {\it JWST}/MIRI observations of SN~2010jl's location at almost 13 yr after explosion.  These observations reveal a bright mid-IR point source coincident with the position of SN~2010jl.  We also obtained late-time visual-wavelength spectra with the {\it MMT} and Keck Observatories, at 7 and 11 yr post-explosion (respectively), which detect emission at the SN location as well.  Our principal conclusions are listed briefly below.

\begin{enumerate}

\item {\it JWST}/MIRI observations reveal a 10--20 $\mu$m source coincident with SN~2010jl that is the brightest MIR point source in the host galaxy.

\item Fits to the MIRI spectrum using various combinations of different grain composition all suggest a total dust mass of at least 0.11--0.3 $M_{\odot}$ has formed by a time of almost 13 yr after explosion.  From the favored optically thick emission + silicate absorption model, we adopt 0.2 $M_{\odot}$ as a conservative estimate of the dust mass at this epoch.  This is among the largest dust masses seen so far in any SN~IIn.  

\item In a scenario where some of the features in the spectrum are caused by absorption by cooler silicate dust, we propose a geometry wherein the cool absorbing dust is located in the inner unshocked SN ejecta, and where the warmer continuum emission arises from dust in the primary CSM interaction region where the forward shock is running through an equatorial ring at the pinched waist of this bipolar shell. Emission from the hot dust on the far side of the shocked equatorial ring must pass through the inner ejecta on its way to us.  Silicate absorption could also potentially occur on the near side of the CSM shell, but because of the absence of significant visual-wavelength absorption at earlier epochs, we favor the interpretation that most of the cooler absorbing dust is in the inner SN ejecta.

\item Late-time visual-wavelength spectra reveal a blue continuum source at the SN position, which is likely to be a young host star cluster.  The spectra also exhibit narrow lines from a surrounding H~{\sc ii} region, and broadened emission lines of H$\alpha$, [O~{\sc i}], and [Ca~{\sc ii}] that correspond to ongoing late-time CSM interaction in SN~2010jl.  These are consistent with a young and very massive progenitor star.

\item The SN emission lines show a pronounced asymmetry, with blueshifted wings extending to roughly  $-$3000 km s$^{-1}$, but with essentially zero flux on the redshifted side of the lines.  This indicates that the similar blueshift seen in IW emission lines in earlier spectra \citep{smith11,smith12,gall14} has persisted to late times when the visual-wavelength continuum of the SN has completely faded. This confirms that the blueshift is due to obscuration by dust within the post-shock CDS, as proposed by \citet{smith12}, and rules out other interpretations involving electron-scattering opacity.

\item The total dust mass of at least 0.11 $M_{\odot}$ almost 13 yr after explosion matches earlier projections for the rate of dust formation in SN~2010jl \citep{gall14}, which were based on comparisons with other SNe such as SN~1987A.  If SN~2010jl continues to follow those projections by forming additional dust in its post-shock CDS or in its ejecta, it may form 1 $M_{\odot}$ in another decade or so.  Whether this dust will be observable in the MIR (as opposed to cooling further and shifting its emission to longer wavelengths) depends on the strength of continued late-time CSM interaction that would be needed to keep it warm. 

\item With much of the dust that formed in SN~2010jl being located in the post-shock CDS, this dust is already behind the shock and is likely to survive, and will contribute to the ISM dust budget.  This is different from the ejecta dust formed in normal SNe, much of which may be destroyed by the reverse shock.

\item Considering the large (and potentially still growing) dust mass seen in SN~2010jl, the common blueshift of emission lines seen in SN~2010jl and other SNe IIn, the location of the dust in the post-shock region (and its likely survival), the large reported  grain size, and other factors, we infer that SNe IIn and SLSNe IIn may make important contributions to the ISM dust budget, despite being $\la$10\% of core-collapse events.  The lower metallicity in the early Universe may even increase their relative contribution if SNe IIn are more common at early epochs, as some indications suggest, but this still remains speculative pending a better understanding of SN~IIn environments.

\end{enumerate}

\begin{acknowledgments}

This work is based on observations made with the NASA/ESA/CSA {\it James Webb Space Telescope}. Support was provided by NASA/{\it JWST} grants GO-01860, GO-3921, AR-06356, and AR-8883. Data were obtained from the Mikulski Archive for Space Telescopes (MAST) at the Space Telescope Science Institute (STScI), which is operated by the Association of Universities for Research in Astronomy, Inc., under NASA contract NAS 5-03127 for JWST. These observations are associated with program \#1860. The specific observations analyzed can be accessed via \dataset[https://doi.org/10.17909/d12d-2n39]{https://doi.org/10.17909/d12d-2n39}.  Support to MAST for these data is provided by the NASA Office of Space Science via grant NAG5–7584 and by other grants and contracts.  Some of the data presented herein were obtained at the W. M. Keck Observatory, which is operated as a scientific partnership among the California Institute of Technology, the University of California, and NASA; the observatory was made possible by the generous financial support of the W. M. Keck Foundation. The authors wish to recognize and acknowledge the very significant cultural role and reverence that the summit of Maunakea has always had within the indigenous Hawaiian community. We are most fortunate to have the opportunity to conduct observations from this mountain.

A.V.F.’s supernova group at UC Berkeley was supported in part by NASA/{\it HST} grants GO-14149 and GO-14668 from STScI, as well as by the
Christopher R. Redlich Fund, Gary and Cynthia Bengier, Clark and Sharon Winslow, Alan Eustace and Kathy Kwan, William Draper, Timothy and Melissa Draper, Briggs and Kathleen Wood,     
Sanford Robertson (W.Z. is a Bengier-Winslow-Eustace Specialist in Astronomy, T.G.B. is a Draper-Wood-Robertson Specialist in Astronomy, Y.Y. was a Bengier-Winslow-Robertson 
Fellow in Astronomy), and numerous other donors.       

\end{acknowledgments}

{\it Facilities:} JWST, HST, Keck, MMT.

\software{Astropy \citep{astropy, astropy:2018, astropy:2022}}

Data Availability.  The JWST raw data associated with this program can be
found at \dataset[DOI:10.17909/d12d-2n39]{https://doi.org/10.17909/d12d-2n39}.

\bibliographystyle{aasjournal}
\bibliography{refs}

\end{document}